\documentclass[a4paper,11pt]{article}

\usepackage{jheppub}

\usepackage{graphicx}
\usepackage{multirow}
\usepackage{bm}
\usepackage{bbm}
\usepackage{color}
\usepackage{slashed}

\title{Novel Method to Reliably Determine the QCD Coupling from $R_{\rm uds}$ Measurements and its effects to Muon $g-2$ and $\alpha(M_Z^2)$ within the Tau-Charm Energy Region}

\author[a]{Jian-Ming Shen,}
\emailAdd{shenjm@hnu.edu.cn}

\author[a]{Bing-Hai Qin,}
\emailAdd{qinbinghai@hnu.edu.cn}

\author[b]{Jiang Yan,}
\emailAdd{yjiang@cqu.edu.cn}

\author[c]{Sheng-Quan Wang}
\emailAdd{sqwang@cqu.edu.cn}

\author[b]{and Xing-Gang Wu}
\emailAdd{wuxg@cqu.edu.cn}

\affiliation[a]{School of Physics and Electronics, Hunan Provincial Key Laboratory of High-Energy Scale Physics and Applications, Hunan University, Changsha 410082, People's Republic of China}
\affiliation[b]{Department of Physics, Chongqing Key Laboratory for Strongly Coupled Physics, Chongqing University, Chongqing 401331, People's Republic of China}
\affiliation[c]{Department of Physics, Guizhou Minzu University, Guiyang 550025, People's Republic of China}

\abstract{We present a novel method for precisely determining the QCD running coupling from $R_{\rm uds}$ measurements in electron-positron annihilation. When calculating the fixed-order perturbative QCD (pQCD) approximant of $R_{\rm uds}$, its effective coupling constant $\alpha_s(Q_*^2)$ is determined by using the principle of maximum conformality, a systematic scale-setting method for gauge theories, whose resultant pQCD series satisfies all the requirements of renormalization group. Contribution due to the uncalculated higher-order (UHO) terms is estimated by using the Bayesian analysis. Using $R_{\rm uds}$ data measured by the KEDR detector at $22$ centre-of-mass energies between $1.84$ GeV and $3.72$ GeV, we obtain $\alpha_s(M_Z^2)=0.1227^{+0.0117}_{-0.0132}({\rm exp.})\pm0.0016({\rm the.})$, where the theoretical uncertainty (the.) is negligible compared to the experimental one (exp.). Numerical analyses confirm that the new method for calculating $R_{\rm uds}$ removes conventional renormalization scale ambiguity, and the residual scale dependence due to the UHO-terms will also be highly suppressed due to a more convergent pQCD series. This leads to a significant stabilization of the perturbative series, and a significant reduction of theoretical uncertainty. It thus provides a reliable theoretical basis for precise determination of the QCD running coupling from $R_{\rm uds}$ measurements at future Tau-Charm Facility. It can also be applied for the precise determination of the hadronic contributions to muon $g-2$ and QED coupling $\alpha(M_Z^2)$ within the tau-charm energy range.}

\begin{document}

\maketitle

\flushbottom

\section{Introduction}

Quantum chromodynamics (QCD) is the fundamental non-Abelian gauge theory of strong interactions. Its running coupling ($\alpha_s$) sets the strength of strong interaction among quarks and gluons, which is crucial and deserves the best possible precision. The strong running coupling becomes weak at short distances due to the property of asymptotic freedom~\cite{Gross:1973id, Politzer:1973fx}, allowing perturbative calculation of physical observables involving large momentum transfer. The strong running coupling in itself is not a physical observable, but rather a quantity defined in the context of perturbation theory, which enters into perturbative QCD (pQCD) predictions for experimentally measurable observables. Its value must be inferred from such measurements and is subject to experimental and theoretical uncertainties~\cite{ParticleDataGroup:2022pth, Deur:2016tte, dEnterria:2022hzv, Deur:2023dzc}.

Total hadronic $e^+e^-$ annihilation rate $R$ is a fundamental observable in QCD, which provides one of the cleanest platforms for determining $\alpha_s$~\cite{Chetyrkin:1996ela}. The $R$ value also contributes to the standard model (SM) prediction for the muon anomalous magnetic moment $a_\mu={(g-2)_\mu}/2$ and the QED running coupling evaluated at the $Z$ pole, $\alpha(M_Z^2)$, e.g., see Refs.\cite{Aoyama:2020ynm, Jegerlehner:2017gek}. Till now, many experimental groups have measured the $R$ value. The recent data \cite{BESIII:2021wib, KEDR:2018hhr} were given by the BES III detector at BEPC II~\cite{BESIII:2009fln} and the KEDR detector at the VEPP-4M $e^+e^-$ collider \cite{Anashin:2013twa}. A collection of all available $R$ data is given in Ref.\cite{ParticleDataGroup:2022pth}. Theoretically, the $R$ value has been evaluated in massless pQCD~\cite{Chetyrkin:1996ela, Davier:2005xq}, and its QCD corrections have now been calculated in the $\overline{\rm MS}$-scheme to order $\alpha_s^4$~\cite{Chetyrkin:1979bj, Gorishnii:1990vf, Surguladze:1990tg, Baikov:2008jh, Baikov:2010je, Baikov:2012er, Baikov:2012zm, Baikov:2012zn}. It has been found that the power suppressed finite-quark-mass effects are well under control~\cite{Chetyrkin:1990kr, Chetyrkin:1994ex, Chetyrkin:1995ii, Chetyrkin:1996cf, Chetyrkin:2000zk, Harlander:2002ur, Kiyo:2009gb} and the same applies to mixed QCD and electroweak corrections \cite{Czarnecki:1996ei, Harlander:1998cmq}.

The $R$ for the continuum light hadron (containing $u$, $d$ and $s$ quarks) production, denoted by $R_{\rm uds}$, is usually adopted to test the validity of pQCD calculation in relatively low energy region~\cite{Kuhn:1998ze, Martin:1999bp}. The measured $R_{\rm uds}(s)$ excludes the contribution from resonances and reflects the lowest order cross section for the inclusive light hadronic event production through one photon annihilation of $e^+e^-$. So it can be directly compared with the pQCD prediction and be directly used to extract $\alpha_s$, or equivalently the QCD scale parameter $\Lambda$.

At present the pQCD calculations for $R_{\rm uds}$ are usually analyzed by using conventional scale-setting method, i.e., one calculates the central value by simply setting the renormalization scale $\mu_r$ equal to the centre-of-mass energy $\mu_r=\sqrt{s}$; and the theoretical uncertainties are estimated by varying the renormalization scale over an arbitrary range such as $\sqrt{s}/2<\mu_r<2\sqrt{s}$. This leads to conventional renormalization scheme and scale ambiguities, and makes the scale uncertainty one of the most important systematic errors for pQCD predictions.

The principle of maximum conformality (PMC)~\cite{Brodsky:2011ta, Brodsky:2011ig, Mojaza:2012mf, Brodsky:2012rj, Brodsky:2013vpa} has been proposed to eliminate the conventional renormalization scale-and-scheme ambiguities. The conventional scale-and-scheme ambiguities are caused by the mismatching of the strong coupling and its corresponding coefficients, since its scale is set by guessing. It is noted that the $\alpha_s$-running behavior is governed by the renormalization group equation (RGE), and then the $\{\beta_i\}$-terms emerged in perturbative series can be inversely adopted for fixing the correct value of $\alpha_s$. The PMC single-scale-setting approach (PMCs)~\cite{Shen:2017pdu, Yan:2022foz} determines an overall effective $\alpha_s$ (its argument is called as the PMC scale) for any fixed order prediction with the help of RGE. The PMC scale can be treated as the effective momentum flow of the process. It has been shown that the PMC prediction is free of conventional scale ambiguity~\cite{Wu:2018cmb, Wu:2019mky}, being consistent with the fundamental renormalization group approaches~\cite{StueckelbergdeBreidenbach:1952pwl,MR0073481,Peterman:1978tb, Wu:2014iba} and the self-consistency requirements of the renormalization group~\cite{Brodsky:2012ms, Wu:2013ei}. However there is still residual scale dependence due to uncalculated higher-order (UHO) terms, which will be highly suppressed due to more convergent pQCD series~\cite{Zheng:2013uja}. The PMC reduces in the Abelian limit to the Gell-Mann-Low method~\cite{Gell-Mann:1954yli} and it provides a systematic way to extend the well-known Brodsky-Lepage-Mackenzie (BLM) method~\cite{Brodsky:1982gc} to all orders.

In this work, we will first adopt the PMCs approach to deal with the perturbative series of $R_{\rm uds}(s)$. Contributions due to the uncalculated higher-order terms will be estimated by using the Bayesian analysis. Then by using the predicted $R_{\rm uds}(s)$ as the basic input, we will extract the value of $\alpha_s$ from the KEDR data on $R_{\rm uds}(s)$ and calculate its effect to Muon $g-2$ and $\alpha(M_Z^2)$.

The remaining parts of the paper are organized as follows. In Sec.~\ref{sec:calctech}, we will present useful formulas for $R_{\rm uds}(s)$ and give a brief description on the calculation procedures. In Sec.~\ref{sec:resudisc}, we will show the numerical results and discussions. Sec.~\ref{sec:summary} is reserved as a summary.

\section{Calculation technology} \label{sec:calctech}

Total hadronic $e^+e^-$ annihilation rate $R(s)$ is related to the theoretically calculable Adler function $D$ as follows~\cite{Adler:1974gd},
\begin{eqnarray}\label{eq:dispersion}
D(Q^2)=-12\pi^2Q^2\frac{d}{d Q^2}\Pi(Q^2)=\int_{4m_\pi^2}^{\infty}\frac{Q^2R(s)ds}{(s+Q^2)^2}.
\end{eqnarray}
Here the Adler function $D$ is defined as the logarithmic derivative of the hadronic vacuum polarization function $\Pi$, which can be written in terms of $\Pi$ and the photon field anomalous dimension, $\gamma$, e.g.~\cite{Baikov:2012zm}
\begin{equation}
D(\alpha_s) = 12\pi^2 \left[\gamma(\alpha_s) - \beta(\alpha_s)\frac{\partial} {\partial{\alpha_s}}{\Pi}(Q^2,{\alpha_s})\right]. \label{eq:Dexpression}
\end{equation}
$\gamma$ and $\Pi$ are given by the perturbative expansions,
\begin{eqnarray}\label{eq:gammaPi}
\gamma=\frac{d_R}{16\pi^2}\sum\limits_{i\geq 0} {\gamma_i}{\left(\frac{\alpha_s}{\pi}\right)^i}, \;\;
\Pi=\frac{d_R}{16\pi^2}\sum\limits_{i\geq 0}{\Pi_i}{\left(\frac{\alpha_s}{\pi}\right)^i}, \nonumber
\end{eqnarray}
where $d_R=N_c$ is the dimension of the quark representation of the colour gauge group. The coefficients $\gamma_i=(\sum_f q_f^2)\gamma_i^{\rm ns}+(\sum_f q_f)^2 \gamma_i^{\rm si}$ and $\Pi_i=(\sum_f q_f^2)\Pi_i^{\rm ns}+(\sum_f q_f)^2 \Pi_i^{\rm si}$, where the superscripts ``ns'' and ``si'' denote the non-singlet and the singlet components, respectively. The singlet contribution starts from order-$\alpha_s^3$, i.e., $\Pi_0^{\rm si}=\Pi_1^{\rm si}=\Pi_2^{\rm si}=0$, $\gamma_0^{\rm si}=\gamma_1^{\rm si}=\gamma_2^{\rm si}=0$. All these perturbative coefficients $\gamma^{\rm ns}_i$, $\gamma^{\rm si}_i$, $\Pi^{\rm ns}_i$ and $\Pi^{\rm si}_i$ up to four-loop QCD corrections can be found in Ref.\cite{Baikov:2012zm}.

Using the perturbative expansions of $\gamma$ and $\Pi$, one then obtains the perturbative expansion for $R(s)$. As for $R_{\rm uds}(s)$, its perturbative expression reads
\begin{eqnarray}\label{eq:Rudsconv}
R_{\rm uds}^{(\ell)}(s) = 2\left[1+\sum_{i=1}^{\ell}r_i \left(\frac{\alpha_s(s)}{\pi}\right)^i\right],
\end{eqnarray}
where $\ell$ specifies the known loop level of the QCD correction, the renormalization scale is set to $\mu_r^2=s$. The results for generic values of $\mu_r$ can be easily recovered by using the standard RGE evolution. The perturbative coefficients $r_i$ can be divided into conformal parts ($r_{i,0}$) and non-conformal parts (proportional to $\beta_i$), i.e. $r_i=r_{i,0}+\mathcal{O}(\{\beta_i\})$. The $\{\beta_i\}$-pattern at different orders exhibits special degeneracies~\cite{Brodsky:2013vpa, Mojaza:2012mf, Bi:2015wea}, which lead to
\begin{eqnarray}
r_1 &=& r_{1,0}, \\
r_2 &=& r_{2,0} + \beta_0 r_{2,1}, \\
r_3 &=& r_{3,0} + \beta_1 r_{2,1} + 2{\beta_0}r_{3,1} + \beta_0^2 r_{3,2}, \\
r_4 &=& r_{4,0} + {\beta_2}{r_{2,1}} + 2{\beta_1}{r_{3,1}} + \frac{5}{2}{\beta_1}{\beta_0}{r_{3,2}} + 3{\beta_0}{r_{4,1}}+ 3\beta_0^2{r_{4,2}} + \beta_0^3{r_{4,3}}, \\
&& \hspace{-4.5mm} \cdots  \nonumber
\end{eqnarray}
where
\begin{eqnarray}\label{eq:rij}
r_{i(\geq 1),0} = {3\over{4}}\gamma^{\rm ns}_i, \;\;\;
r_{i(\geq 2),1} = {3\over{4}}\Pi^{\rm ns}_{i-1}, \;\;\;
r_{i(\geq 3),2} = -{\pi^2 \over {4}}\gamma^{\rm ns}_{i-2}, \;\;\; r_{i(\geq 4),3} = -{3\pi^2\over{4}}\Pi^{\rm ns}_{i-3}.
\end{eqnarray}
It is noted that for $R_{\rm uds}$, only $u$, $d$ and $s$ quarks are produced, thus the number of active flavours is $n_f=3$. Since $\sum_{f=u,d,s} q_f=0$, the singlet contribution vanishes in the present considered three-flavor case. The anomalous dimension $\gamma$ also contains $n_f$ terms, but it governs the QCD-induced corrections to the running of inverse QED coupling constant $\alpha^{-1}$ \cite{Baikov:2012zm}, i.e., ${d (\alpha^{-1})}/{d \ln\mu_r^2}=-\alpha^{-2}{d \alpha}/{d \ln\mu_r^2}=-4\pi\gamma(\alpha_s)$, and is independent to the running of QCD coupling constant, thus its coefficients $\gamma^{\rm ns}_i$ are kept as conformal coefficients that represent the intrinsic perturbative nature of $R(s)$. Starting from $r_3$, terms proportional to $\pi^2$ arise due to continuation of the spacelike perturbative results into the timelike domain. These ``$\pi^2$-terms'' are also called ``kinematical terms'', and can be predicted from those of lower order. It is necessary to emphasize that, Eq.(\ref{eq:Rudsconv}) only partially retains the effects due to continuation of the spacelike perturbative results into the timelike domain, and has certain shortcomings (see, e.g., \cite{Kataev:1995vh, Shirkov:2000qv, Prosperi:2006hx, Nesterenko:2017wpb, Nesterenko:2019rag, Nesterenko:2020nol}). As shown in Eq.(\ref{eq:rij}), all ``$\pi^2$-terms'' are nonconformal, thus will be resummed to a certain level in the PMCs scale-setting procedure.

Following the standard procedure of the PMCs approach~\cite{Shen:2017pdu, Wu:2019mky}, the overall renormalization scale can be determined by requiring all the nonconformal $\{\beta_i\}$-terms vanish, the pQCD approximant (\ref{eq:Rudsconv}) then changes to the following conformal series,
\begin{equation}\label{eq:Rudspmc}
R_{\rm uds}^{(\ell)}(s)|_{\rm PMCs}=2\left[1+\sum_{i=1}^{\ell}r_{i,0}\left(\frac{\alpha_s(Q_*^2)}{\pi}\right)^i\right],
\end{equation}
where the PMC scale $Q_{*}$ is of perturbative nature and can be fixed up to N$^{(\ell-2)}$LL-accuracy, i.e. $\ln Q^2_*/s$ can be expanded as a power series over $\alpha_s(Q_*^2)$,
\begin{equation}\label{eq:scaleformula}
\ln\frac{Q_*^2}{s} = \sum^{\ell-2}_{i=0} S_i \left(\frac{\alpha_s(Q_*^2)}{\pi}\right)^i,
\end{equation}
where the coefficients $S_i\;(i=0,1,2)$ read,
\begin{eqnarray}
S_0 &=& -\frac{\Pi^{\rm ns}_1}{\gamma^{\rm ns}_1}, \label{eq:s0}\\
S_1 &=& \frac{2\gamma^{\rm ns}_2 \Pi^{\rm ns}_1}{{\gamma^{\rm ns}_1}^2}-\frac{2 \Pi^{\rm ns}_2}{\gamma^{\rm ns}_1} +\beta_0 \left(\frac{{\Pi^{\rm ns}_1}^2}{{\gamma^{\rm ns}_1}^2}+\frac{\pi^2}{3}\right), \label{eq:s1}\\
S_2 &=& -\frac{4 {\gamma^{\rm ns}_2}^2 \Pi^{\rm ns}_1}{{\gamma^{\rm ns}_1}^3}+\frac{3 \gamma^{\rm ns}_3 \Pi^{\rm ns}_1}{{\gamma^{\rm ns}_1}^2}
   +\frac{4 \gamma^{\rm ns}_2 \Pi^{\rm ns}_2}{{\gamma^{\rm ns}_1}^2}-\frac{3 \Pi^{\rm ns}_3}{\gamma^{\rm ns}_1} \nonumber\\
&& +\beta_0 \left(\frac{\pi^2 \gamma^{\rm ns}_2}{3 \gamma^{\rm ns}_1}-\frac{5 \gamma^{\rm ns}_2 {\Pi^{\rm ns}_1}^2}{{\gamma^{\rm ns}_1}^3}+\frac{6 \Pi^{\rm ns}_1 \Pi^{\rm ns}_2}{{\gamma^{\rm ns}_1}^2}\right)
+\beta_1\left(\frac{3{\Pi^{\rm ns}_1}^2}{2{\gamma^{\rm ns}_1}^2}+\frac{\pi^2}{2}\right) -\beta_0^2\frac{2{\Pi^{\rm ns}_1}^3}{{\gamma^{\rm ns}_1}^3}.
\end{eqnarray}
Eq.(\ref{eq:scaleformula}) shows that the logarithmic form $\ln Q^2_*/s$ is a power series in $\alpha_s$, which resums all the known $\{\beta_i\}$-terms via the RGE, and is independent of $\mu_r$ at any fixed order.

The resulting conformal series (\ref{eq:Rudspmc}) with an overall $\alpha_s(Q_*)$ provides not only precise prediction for the known fixed-order pQCD series, but also a reliable basis for estimating the contributions from the uncalculated higher-order (UHO) terms. As an estimation of the UHO terms of the perturbative series, we adopt a Bayesian-based approach (BA) \cite{Cacciari:2011ze, Shen:2022nyr} to quantify it in terms of a probability distribution. The conditional probability density function (p.d.f.) $f_c(c_{n}|c_l,c_{l+1},\dots,c_k)$ for a generic (uncalculated) coefficient $c_{n}$ ($n>k$) of any possible perturbative series $\rho_k=\sum_{i=l}^{k}c_i\alpha_s^i$ with given coefficients $\{c_l,c_{l+1},\dots,c_k\}$ is given by
\begin{eqnarray}
\label{eq:conditionalpdf1}
f_c(c_n|c_l,\cdots,c_k)=\int h_0(c_n|\bar c)f_{\bar c}(\bar c|c_l,\cdots,c_k) {\rm d}{\bar c},\;\;
\end{eqnarray}
where ${\bar c}$ ($>0$) is a common boundary for the absolute values of all the known coefficients $\{c_l,\dots,c_k\}$ and the unknown coefficient $c_n$ one wants to evaluate. $h_0(c_n|\bar c)$ is the conditional p.d.f. of $c_n$ given ${\bar c}$. The conditional p.d.f. of ${\bar c}$ given coefficients $\{c_l,\cdots,c_k\}$, $f_{\bar c}({\bar c}|c_l,\cdots,c_k)$, can be determined by applying the Bayes' theorem,
\begin{eqnarray}
\label{eq:bayes}
f_{\bar c}(\bar c|c_l,\cdots,c_k)=\frac{h(c_l,\cdots,c_k|\bar c)g_0(\bar c)}{\int h(c_l,\cdots,c_k|\bar c)g_0(\bar c) {\rm d}{\bar c}}\;,
\end{eqnarray}
where $h(c_l,\cdots,c_k|\bar c)$ is the \emph{likelihood function} for $\bar{c}$ ; i.e., the joint p.d.f. for the coefficients viewed as a function of $\bar{c}$, evaluated with coefficients actually obtained in the calculation. The function $g_0(\bar c)$ is the prior p.d.f. for ${\bar c}$. Both $g_0(\bar c)$ and $h_0(c_i|\bar c)$ depend on the model assumption. Here we use the CH model \cite{Cacciari:2011ze}, which suggests: both $\ln{\bar c}$ and $c_i$ are equally probable for all their possible values; all the coefficients that we know and that we want to evaluate are mutually independent with the exception for the common bound (${\bar c}$), which results in $h(c_l,\cdots,c_k|{\bar c})=\prod_{i=l}^k h_0(c_i|{\bar c})$. Using the CH model, we obtain a symmetric posterior distribution for negative and positive $c_n$: a central plateau with suppressed tails \cite{Cacciari:2011ze, Shen:2022nyr}. The knowledge of p.d.f. $f_c(c_{n}|c_l,c_{l+1},\dots,c_k)$ allows one to calculate the degree-of-belief (DoB) that the value of $c_{n}$ belongs to some credible interval (CI). The symmetric smallest CI of fixed $p\%$ DoB for $c_{n}$ is,
\begin{eqnarray}\label{eq:CI}
c_n\in[-c_n^{(p)},c_n^{(p)}]\;,
\end{eqnarray}
where the boundary $c_n^{(p)}$ is defined implicitly by
\begin{eqnarray}\label{eq:DoB}
p\% = \int_{-c_n^{(p)}}^{c_n^{(p)}} f_c(c_n |c_l,\dots,c_k) {\rm d} c_n.
\end{eqnarray}
The expression of $c_n^{(p)}$ can be found in Ref.\cite{Shen:2022nyr}. We take $p\%=95.5\%$ in the following calculation.

\section{Numerical results and discussions}
\label{sec:resudisc}

In numerical calculation, to be consistent we shall adopt the $\ell$-loop $\alpha_s$-running to obtain numerical predictions for $R_{\rm uds}^{(\ell)}(s)$.

\subsection{Basic properties of the pQCD approximant for $R_{\rm uds}(s)$.}

The PMC prediction for $R_{\rm uds}(s)$ up to order $\alpha_s^4$ reads,
\begin{eqnarray}\label{eq:pmcs}
\frac{1}{2}R_{\rm uds}^{(4)}(s)|_{\rm PMCs}=1+\frac{\alpha_s(Q_*^2)}{\pi}+0.2174\alpha_s^2(Q_*^2)
+0.1108\alpha_s^3(Q_*^2)+0.0698\alpha_s^4(Q_*^2),
\end{eqnarray}
where $Q_*$ can be fixed up to N$^2$LL accuracy,
\begin{eqnarray}\label{eq:pmcscale}
\ln\frac{Q_*^2}{s}=0.2249+1.5427\alpha_s(Q_*^2)+2.4933\alpha_s^2(Q_*^2).
\end{eqnarray}
Both the PMC conformal series (\ref{eq:pmcs}) and the PMC scale (\ref{eq:pmcscale}) are scale-independent, which will have residual scale dependence due to uncalculated terms~\cite{Zheng:2013uja}.

As a comparison, we present the conventional prediction for $R_{\rm uds}(s)$ by taking $\mu_r=\sqrt{s}$,
\begin{eqnarray}\label{eq:conv}
\frac{1}{2}R_{\rm uds}^{(4)}(s)|_{\rm Conv.}=1+\frac{\alpha_s(s)}{\pi}+0.1661\alpha_s^2(s)-0.3317\alpha_s^3(s)-1.0972\alpha_s^4(s).
\end{eqnarray}

The UHO coefficients predicted by using BA are $r_{5,0}/\pi^5\in[-0.4622,0.4622]$ for the pQCD approximant (\ref{eq:pmcs}) and $S_3/\pi^3\in[-4.4159,4.4159]$ for the PMC scale (\ref{eq:pmcscale}). More coefficients at lower order predicted by using BA are given in Tables \ref{tab:Qstarcoe} and \ref{tab:Rudsri0}, where the conventional coefficients $r_i$ with fixed $\mu_r=\sqrt{s}$ are also presented as a comparison. It is noted that almost all the exact values of these coefficients lie within the predicted $95.5\%$ credible intervals. There are only one exception for the conventional coefficient $r_4(\mu_r=\sqrt{s})$, i.e., the exact value of $r_4(\mu_r=\sqrt{s})$ is outside the region of the $95.5\%$ credible interval predicted by using the BA based on the known coefficients $r_i(\mu_r=\sqrt{s})$ ($i=1,2,3$).

Because the known coefficients of the conventional pQCD series (\ref{eq:conv}) are scale-dependent at every order,
the BA can only be applied after one specifies the choices for the renormalization scale, thus introducing extra uncertainties for the BA. Such extra uncertainty can be simply evaluated by varying the renormalization scale $\mu_r$ in some range, such as, $\sqrt{s}/2<\mu_r<2\sqrt{s}$. This variation range will be labelled as $\Delta\mu_r$ in the following.
There are improved models based on the Bayesian analysis, i.e., the geometric model \cite{Bonvini:2020xeo} and the abc model \cite{Duhr:2021mfd}, which can be applied to deal with the conventional pQCD series with conventional scale dependence $\Delta\mu_r$, or the PMC series with residual scale dependence $\Delta Q_*$. We stress that in Refs. \cite{Bonvini:2020xeo, Duhr:2021mfd} ways to deal with the scale dependence within the Bayesian approach have been introduced, and further investigation is left to future work.


\begin{table}[htbp]
\centering
\caption{The predicted $95.5\%$ credible intervals (CI) for the UHO coefficients of $\ln{Q_*^2}/{s}$ via the Bayesian-based approach. The exact values (EV) are presented as a comparison.}
\begin{tabular}[b]{cccc}
\hline
 & ~$S_1/\pi$~ & ~$S_2/\pi^2$~ & ~~$S_3/\pi^3$~~  \\
\hline
CI & $[-2.4988,2.4988]$ & $[-4.1986,4.1986]$ & $[-4.4159,4.4159]$ \\
EV & $1.5427$ & $2.4933$ & - \\
\hline
\end{tabular}
\label{tab:Qstarcoe}
\end{table}

\begin{table}[htbp]
\centering
\caption{The predicted $95.5\%$ credible intervals (CI) for the scale-dependent conventional coefficients $r_i(\mu_r)$ ($i=2,3,4,5$) at fixed $\mu_r=\sqrt{s}$ and the scale-invariant coefficients $r_{i,0} (i=2,3,4,5)$ of $R_{\rm uds}^{(\ell)}(s)$ via the Bayesian-based approach. The exact values (EV) are presented as a comparison.}
\begin{tabular}[b]{ccccc}
\hline
 & ~$r_{2,0}/\pi^2$~ & ~$r_{3,0}/\pi^3$~ & ~~$r_{4,0}/\pi^4$~~ & ~~$r_{5,0}/\pi^5$~~ \\
\hline
CI & $[-3.5368,3.5368]$ & $[-0.8663,0.8663]$ & $[-0.5638,0.5638]$ & $[-0.4622,0.4622]$ \\
EV & $0.2174$ & $0.1108$ & $0.0698$ & - \\
\hline
 & ~$r_2(\mu_r=\sqrt{s})/\pi^2$~ & ~$r_3(\mu_r=\sqrt{s})/\pi^3$~ & ~$r_4(\mu_r=\sqrt{s})/\pi^4$~ & ~$r_5(\mu_r=\sqrt{s})/\pi^5$~ \\
\hline
CI & $[-3.5368,3.5368]$ & $[-0.8663,0.8663]$ & $[-0.5874,0.5874]$ & $[-1.5931,1.5931]$ \\
EV & $0.1661$ & $-0.3317$ & $-1.0972$ & - \\
\hline
\end{tabular}
\label{tab:Rudsri0}
\end{table}

\begin{table}[htbp]
\centering
\caption{The PMC scale $Q_*$ at various orders with different input $\sqrt{s}/\Lambda=5,8,12$, respectively. The central values are calculated according to Eq.(\ref{eq:pmcscale}) truncated at corresponding accuracy, and the uncertainties are estimated by using BA.} \label{tab:scale}
\begin{tabular}{c|lccc}
\hline
~Input~ & & ~~LL~~ & ~~NLL~~ & ~~N$^{2}$LL~~ \\  \hline
~$\sqrt{s}/\Lambda=5$~ & ~$Q_*/\Lambda$~ & $5.60^{+2.05}_{-2.25}$ & $6.92^{+0.95}_{-1.07}$ & $7.50^{+0.26}_{-0.27}$ \\  \hline
~$\sqrt{s}/\Lambda=8$~ & ~$Q_*/\Lambda$~ & $8.95^{+2.71}_{-2.58}$ & $10.68^{+1.08}_{-1.13}$ & $11.33^{+0.25}_{-0.26}$ \\  \hline
~$\sqrt{s}/\Lambda=12$~ & ~$Q_*/\Lambda$~ & $13.43^{+3.53}_{-3.26}$ & $15.67^{+1.24}_{-1.27}$ & $16.41^{+0.25}_{-0.26}$ \\  \hline
\end{tabular}
\end{table}

\begin{figure}[htbp]
\centering
\includegraphics[width=0.6\textwidth]{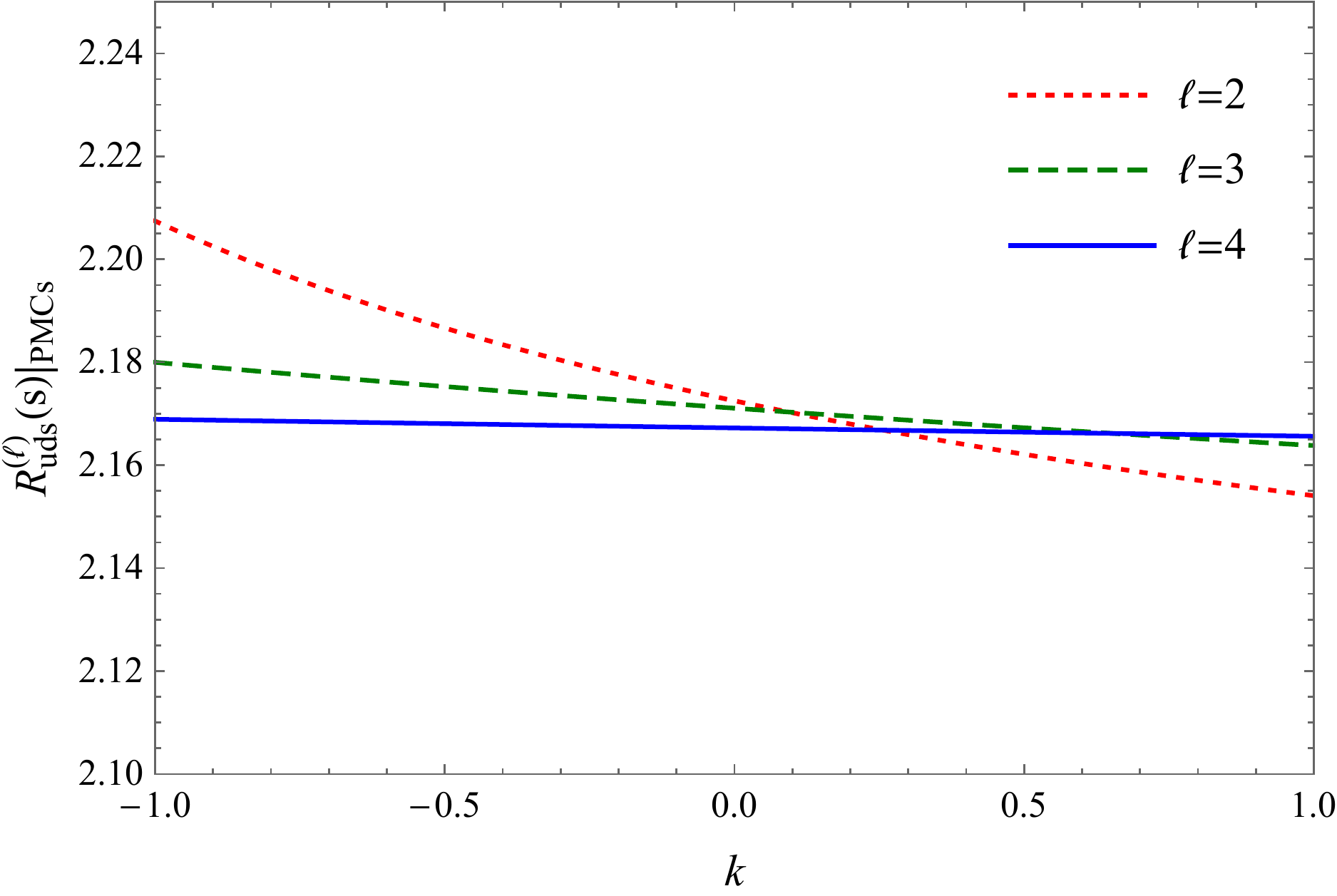}
\caption{The residual scale dependence of the PMC prediction $R_{\rm uds}^{(\ell)}(s)|_{\rm PMCs}$ ($\ell=2,3,4$) at fixed $\sqrt{s}/\Lambda=8$. The dotted, dashed, and solid lines represent $R_{\rm uds}(s)$ up to $2$-loop, $3$-loop, and $4$-loop, respectively. $k\in [-1,1]$ is defined by $Q_*=Q_{*,{\rm central}}\pm k|\Delta Q_*|_{\rm MAX}$.}
\label{fig:residual}
\end{figure}

\begin{figure}[htbp]
\centering
\includegraphics[width=0.6\textwidth]{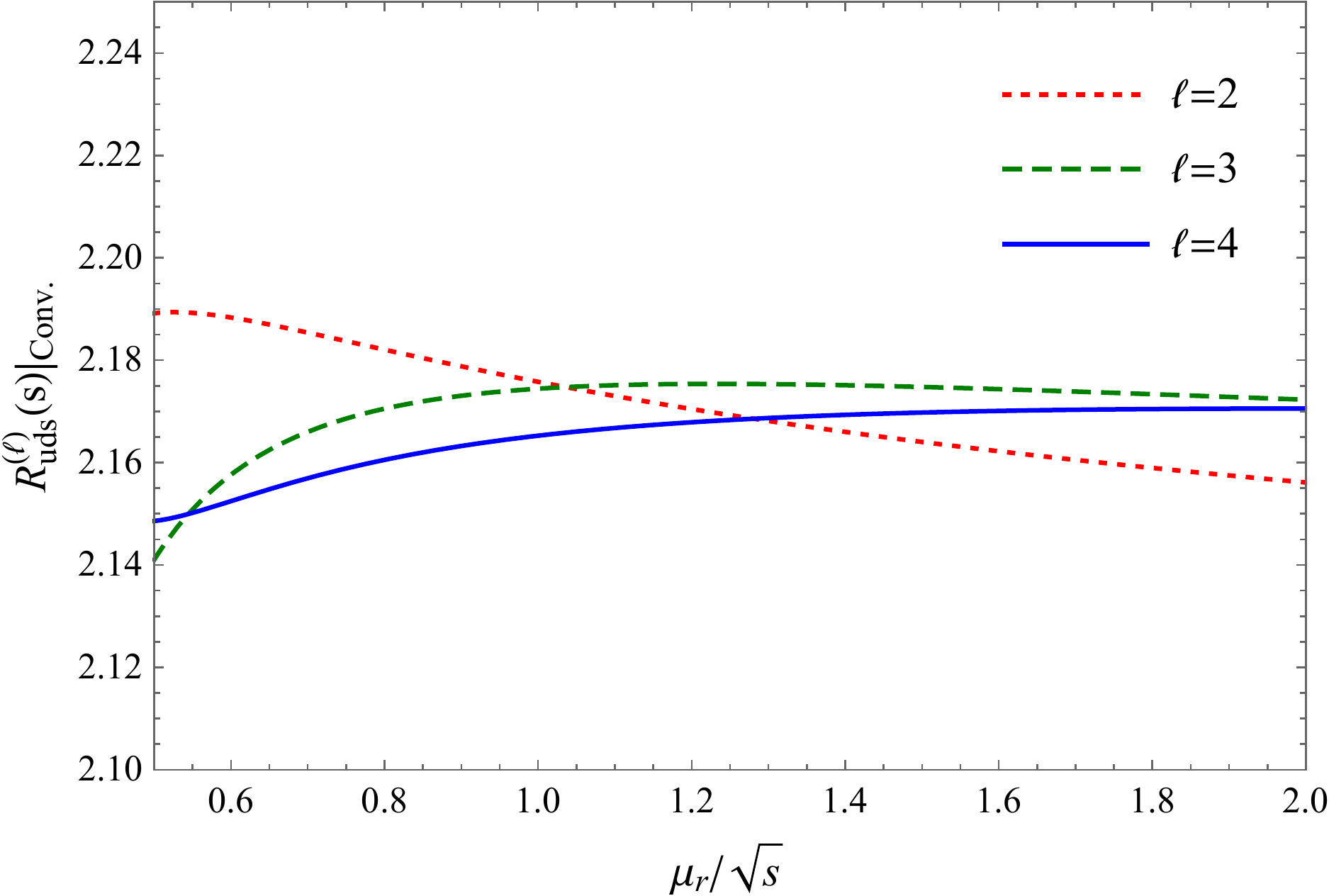}
\caption{The conventional scale dependence of the conventional prediction $R_{\rm uds}^{(\ell)}(s)|_{\rm Conv.}$ ($\ell=2,3,4$) at fixed $\sqrt{s}/\Lambda=8$. The dotted, dashed, and solid lines represent $R_{\rm uds}(s)$ up to $2$-loop, $3$-loop, and $4$-loop, respectively.}
\label{fig:RudsVSscale}
\end{figure}

Firstly, we present the calculated PMC scales at various orders with three different input $\sqrt{s}/\Lambda=5,8,12$ in Table \ref{tab:scale}. Here and following, unless otherwise specified, $\Lambda=\Lambda_{\overline{\rm MS}}^{(3)}$, represents the $\Lambda$ parameter of $\overline{\rm MS}$ scheme in three-flavor QCD. In Table \ref{tab:scale}, the central values are calculated according to Eq.(\ref{eq:pmcscale}) truncated at corresponding accuracy, and the errors $\Delta Q_*$ are determined by taking the UHO coefficients of $\ln{Q_*^2}/{s}$ presented in Table \ref{tab:Qstarcoe}. To show the residual scale dependence of the PMC predictions, we present $R_{\rm uds}^{(\ell)}(s)|_{\rm PMCs}$ ($\ell=2,3,4$) as a function of $k$ with fixed $\sqrt{s}/\Lambda=8$ in Figure \ref{fig:residual}, where $k\in [-1,1]$ is defined by $Q_*=Q_{*,{\rm central}}\pm k|\Delta Q_*|_{\rm MAX}$ with $Q_{*,{\rm central}}$ the central value and $|\Delta Q_*|_{\rm MAX}$ the maximum of the absolute values of lower and upper errors. As a comparison, the conventional scale dependence of the conventional predictions $R_{\rm uds}^{(\ell)}(s)|_{\rm Conv.}$ ($\ell=2,3,4$) is presented in Figure \ref{fig:RudsVSscale}. Figure \ref{fig:residual} shows a reduction of the residual scale dependence for the PMC predictions when increasing the order. While the conventional scale dependence of the conventional predictions is moderate when more-and-more loop corrections have been added as show by Figure \ref{fig:RudsVSscale}.

\begin{figure}[htbp]
\centering
\includegraphics[width=0.6\textwidth]{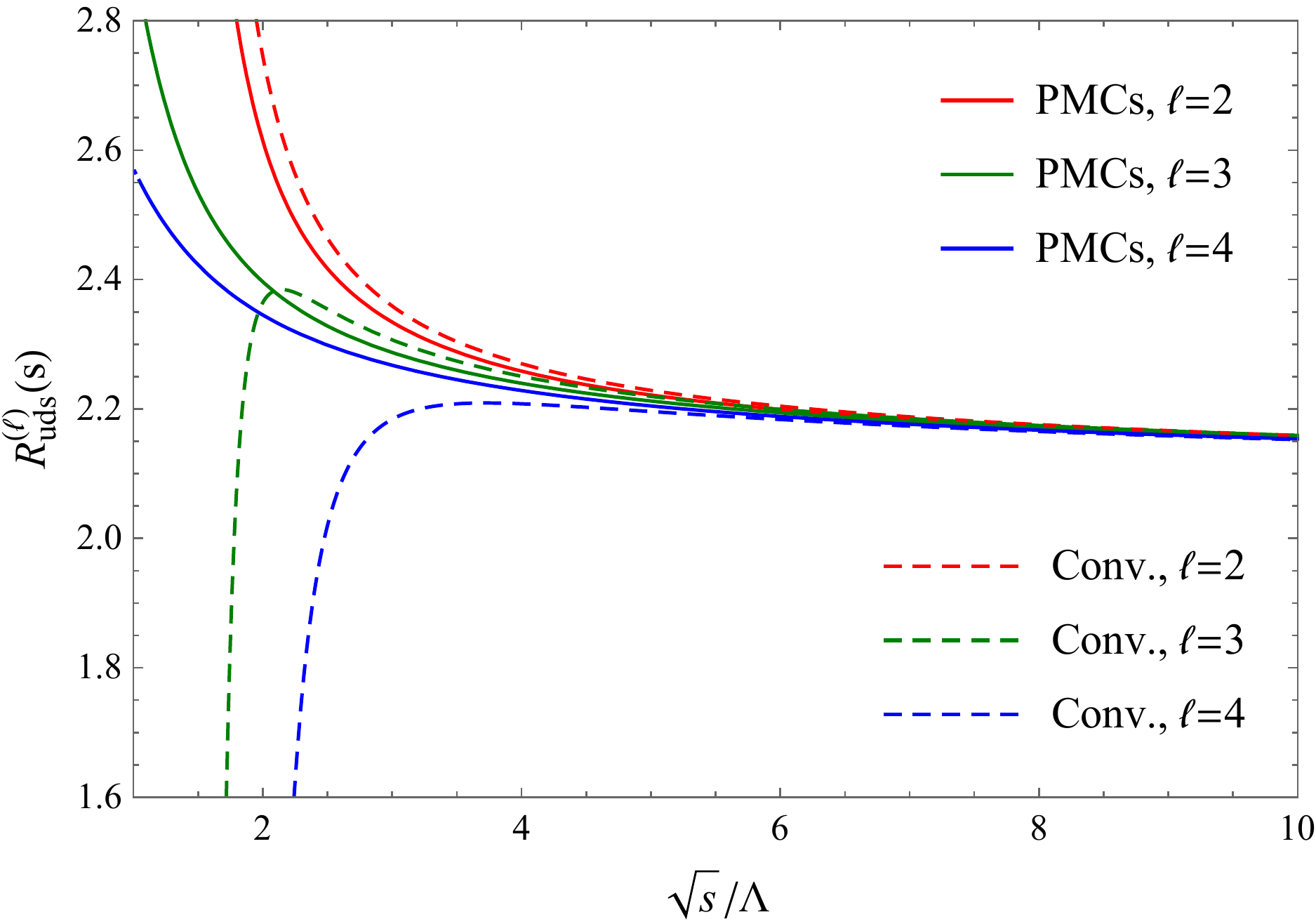}
\caption{$R^{(\ell)}_{\rm uds}(s)$ as a function of $\sqrt{s}/\Lambda$ with $\ell$-loop ($\ell=2,3,4$) QCD corrections. The red, green and blue dashed curves are for conventional predictions by taking $\mu_r\equiv \sqrt{s}$ at $2$-loop, $3$-loop, and $4$-loop, respectively. The red, green and blue thin curves are for PMCs predictions at $2$-loop, $3$-loop, and $4$-loop, respectively.}
\label{fig:Ruds}
\end{figure}

\begin{figure}[htbp]
\centering
\includegraphics[width=0.6\textwidth]{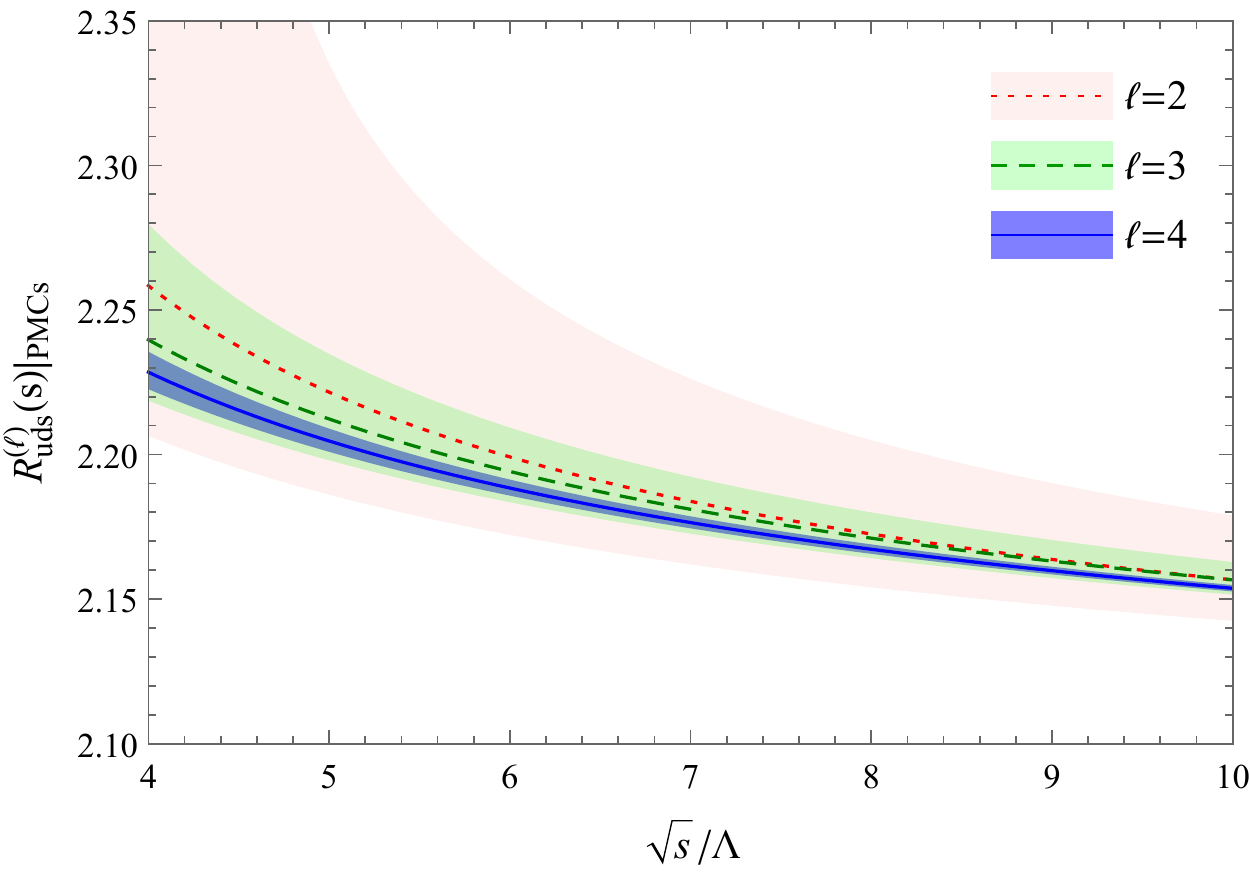}
\caption{The PMCs prediction $R^{(\ell)}_{\rm uds}(s)|_{\rm PMCs}$ ($\ell=2,3,4$) as a function of $\sqrt{s}/\Lambda$. The red dotted, green dashed and blue solid curves are for the predictions at $2$-loop, $3$-loop, and $4$-loop, respectively. For each curve, its error band represents the uncertainty from the residual scale dependence.}
\label{fig:Rudsband}
\end{figure}

\begin{figure}[htbp]
\centering
\includegraphics[width=0.6\textwidth]{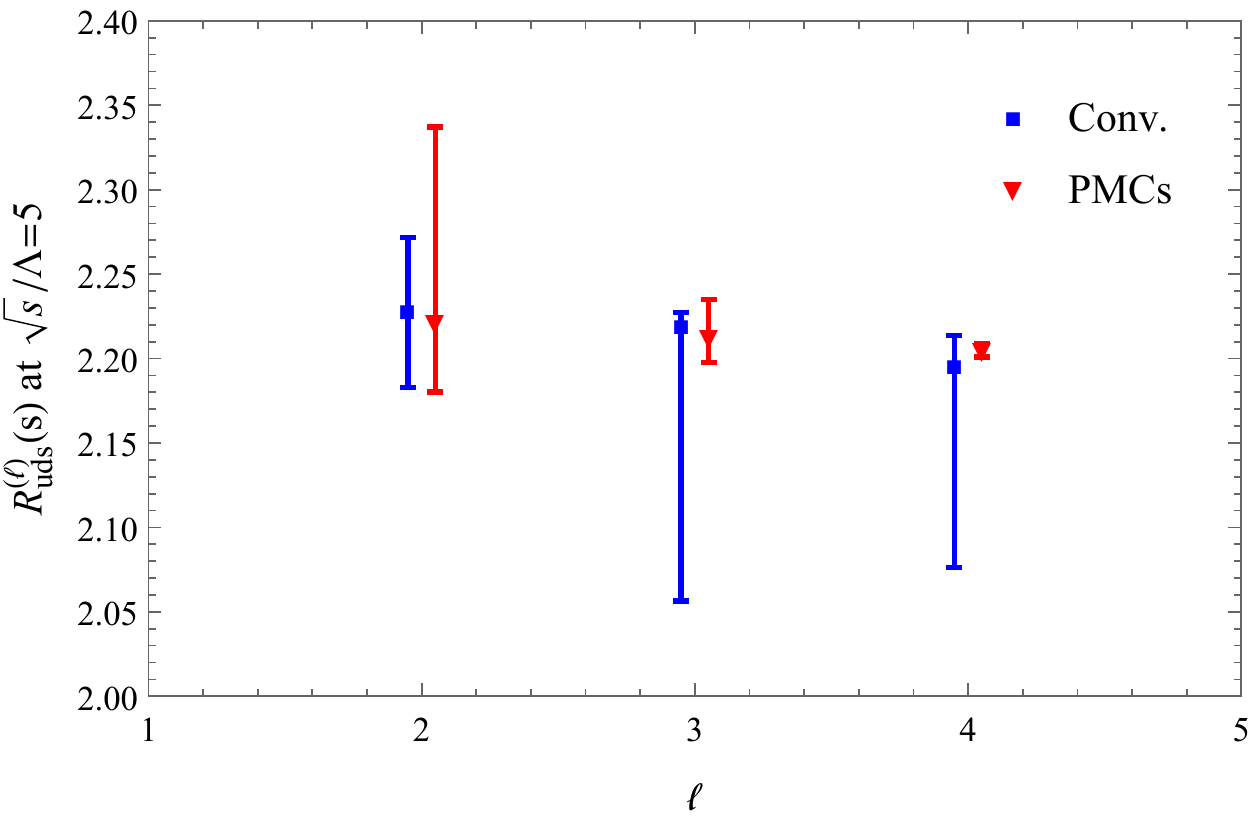}
\includegraphics[width=0.6\textwidth]{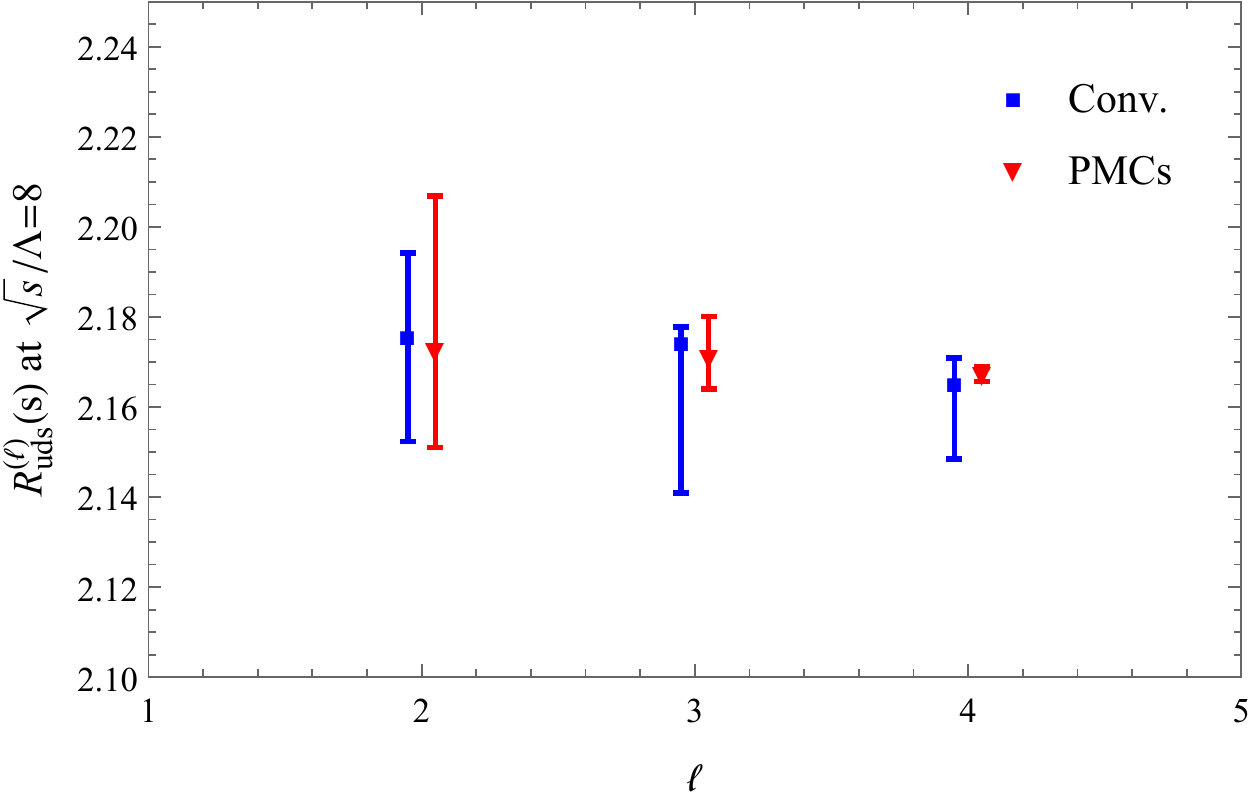}
\includegraphics[width=0.6\textwidth]{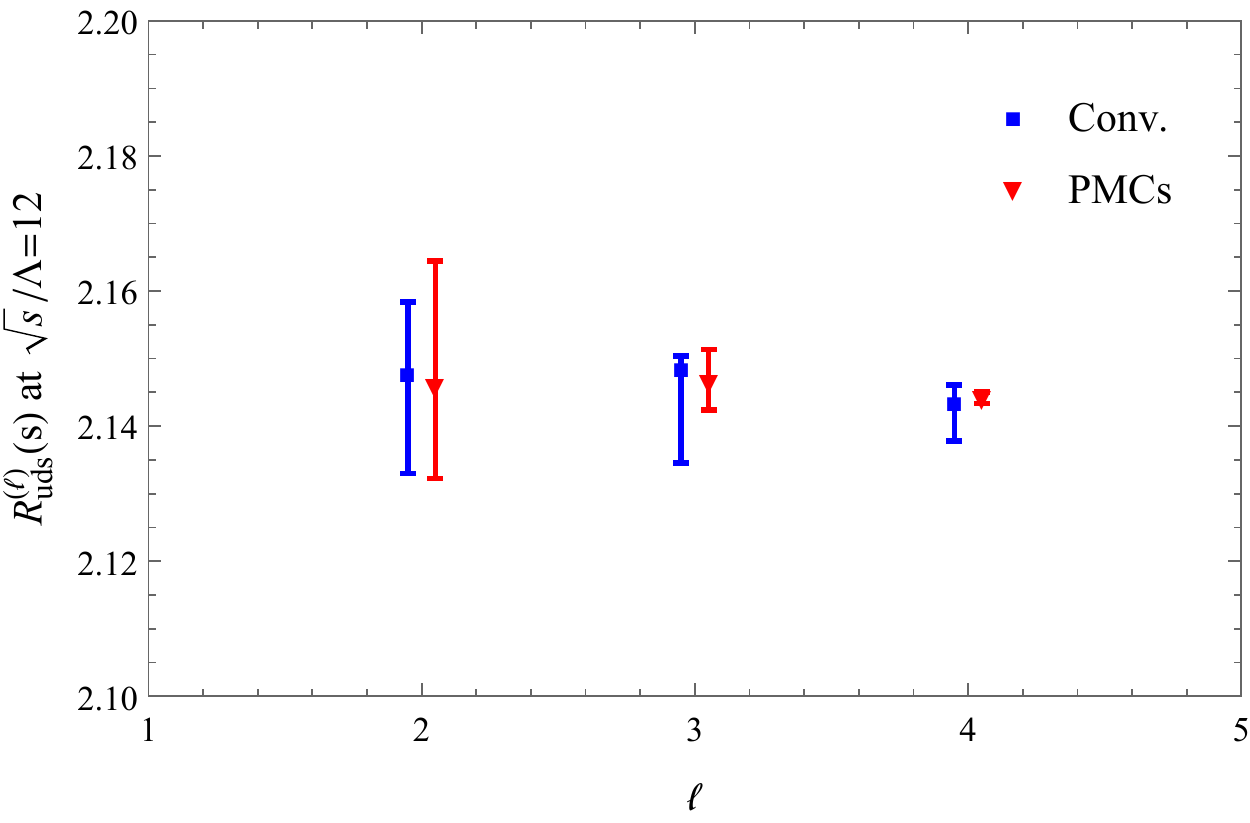}
\caption{Comparison of $R^{(\ell)}_{\rm uds}(s)$ at $\ell$-loops $(\ell=2,3,4)$.  From up to down, the subgraphs are for $\sqrt{s}/\Lambda=5,8,12$, respectively. The blue solid squares and red solid triangles represent the conventional (Conv.) and PMCs predictions, respectively. The error bars include those uncertainties listed in Table \ref{tab:Rn} added in quadrature.}
\label{fig:Rn}
\end{figure}

Secondly, we present the centre-of-mass energy dependence of $R^{(\ell)}_{\rm uds}(s)$ with various loop ($\ell=2,3,4$) QCD corrections in Figure \ref{fig:Ruds}. The red, green and blue dashed curves are for conventional predictions by fixing $\mu_r\equiv \sqrt{s}$ at $2$-loop, $3$-loop, and $4$-loop, respectively. The red, green and blue thin curves are for PMCs predictions at $2$-loop, $3$-loop, and $4$-loop, respectively. As shown by Figure \ref{fig:Ruds}, the loop convergence of $R_{\rm uds}^{(\ell)}(s)$ has been markedly improved after applying the PMCs scale-setting procedure. The conventional prediction (\ref{eq:conv}), whose validity range has been demonstrated to be strictly limited to $\sqrt{s}/\Lambda>{\rm exp}(\pi/2)\simeq4.81$ \cite{Nesterenko:2017wpb, Nesterenko:2019rag, Nesterenko:2020nol}, converges rather slowly when the centre-of-mass energy $\sqrt{s}$ approaches this value, and the corresponding curves start to swerve quite above the boundary of its convergence range. A partial enlarged description for the PMCs predictions $R^{(\ell)}_{\rm uds}(s)|_{\rm PMCs}$ with $\ell$-loop ($\ell=2,3,4$) QCD corrections is presented in Figure \ref{fig:Rudsband}, where the error band of each curve represents the uncertainty from the residual scale dependence. For definiteness, we present the numerical results of $R_{\rm uds}^{(\ell)}(s)$ by taking various QCD corrections ($\ell=2,3,4$)  with three different input $\sqrt{s}/\Lambda=5,8,12$ in Table \ref{tab:Rn}, where the first errors are from the conventional scale dependence $\Delta\mu_r$ of conventional predictions (Conv.), or the residual scale dependence $\Delta Q_*$ of PMCs predictions (PMCs), and the second errors are from the UHO of the pQCD approximant $R_{\rm uds}^{(\ell)}(s)$. Table \ref{tab:Rn} shows that, the theoretical uncertainty of $R_{\rm uds}^{(\ell)}(s)$ is dominated by the scale error, i.e., the conventional $\Delta\mu_r$ for conventional predictions, or the residual $\Delta Q_*$ for PMCs predictions. Note that the uncertainty due to non-perturbative contribution is tiny compared to the scale error as we will show in the following discussions. The errors from $\Delta Q_*$ and the UHO contribution decrease rapidly as the order increases. This has been shown more clearly in Figure \ref{fig:Rn}.

\begin{table}[htbp]
\centering
\caption{$R_{\rm uds}^{(\ell)}(s)$ ($\ell=2,3,4$) with different input $\sqrt{s}/\Lambda=5,8,12$, respectively. The first errors are for $\Delta\mu_r$ of conventional predictions (Conv.), or $\Delta Q_*$ of PMCs predictions (PMCs), the second errors are for UHO of the pQCD approximant $R_{\rm uds}^{(\ell)}(s)$.} \label{tab:Rn}
\begin{tabular}{lcccc}
\hline
& Input & $R_{\rm uds}^{(2)}(s)$ & $R_{\rm uds}^{(3)}(s)$ & $R_{\rm uds}^{(4)}(s)$ \\  \hline\hline
Conv. & $\sqrt{s}/\Lambda=5$ & $2.2287^{+0.0347+0.0256}_{-0.0380-0.0256}$ & $2.2194^{-0.0043+0.0066}_{-0.1627-0.0066}$ & $2.1961^{+0.0164+0.0061}_{-0.1197-0.0061}$ \\  \hline
PMCs & $\sqrt{s}/\Lambda=5$ & $2.2216^{+0.1137+0.0212}_{-0.0355-0.0212}$ & $2.2123^{+0.0225+0.0032}_{-0.0143-0.0032}$ & $2.2047^{+0.0043+0.0006}_{-0.0037-0.0006}$ \\  \hline\hline
Conv. & $\sqrt{s}/\Lambda=8$ & $2.1758^{+0.0134+0.0127}_{-0.0197-0.0127}$ & $2.1744^{-0.0022+0.0026}_{-0.0334-0.0026}$ & $2.1653^{+0.0053+0.0018}_{-0.0167-0.0018}$ \\  \hline
PMCs & $\sqrt{s}/\Lambda=8$ & $2.1725^{+0.0325+0.0111}_{-0.0184-0.0111}$ & $2.1711^{+0.0089+0.0015}_{-0.0069-0.0015}$ & $2.1672^{+0.0017+0.0003}_{-0.0017-0.0003}$ \\  \hline\hline
Conv. & $\sqrt{s}/\Lambda=12$ & $2.1478^{+0.0069+0.0079}_{-0.0125-0.0079}$ & $2.1485^{-0.0013+0.0014}_{-0.0139-0.0014}$ & $2.1435^{+0.0025+0.0008}_{-0.0057-0.0008}$ \\  \hline
PMCs & $\sqrt{s}/\Lambda=12$ & $2.1459^{+0.0171+0.0071}_{-0.0117-0.0071}$ & $2.1465^{+0.0047+0.0009}_{-0.0040-0.0009}$ & $2.1442^{+0.0008+0.0001}_{-0.0008-0.0001}$ \\  \hline\hline
\end{tabular}
\end{table}

\begin{table}[htbp]
\centering
\caption{Total and individual-order QCD corrections for $R^{(4)}_{\rm uds}(s)$ with fixed $\sqrt{s}/\Lambda=8$. The errors of PMCs predictions are for the residual scale dependence, $\Delta Q_*$. The central values of conventional predictions (Conv.) are for $\mu_{r}=\sqrt{s}$, and the errors are for $\Delta \mu_r$.} \label{tab:Ruds4}
\begin{tabular}{lccccc}
\hline
 & NLO & N$^{2}$LO & N$^{3}$LO & N$^{4}$LO & Total \\  \hline
Conv. & $0.1643^{-0.0378}_{+0.0791}$ & $0.0221^{+0.0159}_{-0.0659}$ & $-0.0114^{+0.0191}_{-0.0475}$ & $-0.0097^{+0.0081}_{+0.0176}$ & $0.1653^{+0.0053}_{-0.0167}$ \\  \hline
PMCs & $0.1426_{-0.0012}^{+0.0012}$ & $0.0218_{-0.0004}^{+0.0004}$ & $0.0025_{-0.0001}^{+0.0001}$ & $0.0003_{-0.0000}^{+0.0000}$ & $0.1672_{-0.0017}^{+0.0017}$\\  \hline
\end{tabular}
\end{table}

Thirdly, we present the values of individual-order QCD correction terms for the four-loop predictions $R^{(4)}_{\rm uds}(s)$ with fixed $\sqrt{s}/\Lambda=8$ in Table \ref{tab:Ruds4}. The relative importance among the NLO-terms, the N$^2$LO-terms, the N$^3$LO-terms and the N$^4$LO-terms are
\begin{eqnarray}
&& 1:+0.1345:-0.0694:-0.0590, \;\;\; ({\rm Conv.}, \sqrt{s}/\Lambda=8, \mu_r=\sqrt{s}), \\
&& 1:+0.1529:+0.0175:+0.0021, \;\;\; ({\rm PMCs}, \sqrt{s}/\Lambda=8).
\end{eqnarray}
These two equations show that improved convergence can be obtained after using the PMCs scale-setting procedure. Table \ref{tab:Ruds4} shows that there are residual scale dependence for every order of the PMCs prediction, which, however, are markedly smaller than the conventional scale dependence of the conventional prediction. Note that with fixed $\sqrt{s}/\Lambda=8$, the scale variation in $\sqrt{s}/2<\mu_r<2\sqrt{s}$ leads to $\sim10.1\%$ variation for the QCD correction of $R_{\rm uds}^{(4)}(s)$, and $\sim0.8\%$ variation for the whole four-loop prediction $R_{\rm uds}^{(4)}(s)$. The PMC series, which has smaller (residual) scale dependence and a good convergent behavior, can be treated as the intrinsic perturbative nature of the series.

It should be pointed out that there are extra ``power-correction'' ${\cal O}(1/Q^4)$ to the total cross section in $e^+e^-$ annihilation, which accounts for contributions that are fundamentally non-perturbative \cite{Braaten:1991qm, Dokshitzer:1995qm}. These are introduced via non-vanishing vacuum expectation values originating from quark and gluon condensation. The non-perturbative addition to the Adler function has been calculated~\cite{Braaten:1991qm, Davier:1997vd},
\begin{eqnarray}
\label{eq:nonpdef}
D_{\rm NP}(-s) &=& N_C \sum_f Q_f^2
\Bigg\{ \frac{2\pi^2}{3}\left(1 - \frac{11}{18}\frac{\alpha_s(s)}{\pi} \right)\frac{\left\langle\frac{\alpha_s}{\pi} GG\right\rangle}{s^2} + 8\pi^2\left(1 - \frac{\alpha_s(s)}{\pi}\right)\frac{\langle m_f\bar{q_f}q_f\rangle}{s^2}  \nonumber \\
& &\hspace{2.2cm}
+  \frac{32\pi^2}{27}\frac{\alpha_s(s)}{\pi}
\sum_k\frac{\langle m_k\bar{q_k}q_k\rangle}{s^2}
+ 12\pi^2\frac{\langle{\cal O}_6\rangle}{s^3}
+ 16\pi^2\frac{\langle{\cal O}_8\rangle}{s^4} \Bigg\}~,
\end{eqnarray}
where the non-perturbative operators are the gluon condensate, $\langle(\alpha_s/\pi) GG\rangle$, and the quark condensates, $\langle m_f\bar{q_f}q_f\rangle$. The latter obey approximately the partially conserved axial-vector current relations \cite{Gell-Mann:1960mvl, Nambu:1960xd, Davier:1998si},
\begin{eqnarray}
\label{eq_dnp}
(m_u + m_d)\langle\bar{u}u + \bar{d}d\rangle \simeq  - 2 f_\pi^2 m_\pi^2~, \;\;\; m_s\langle\bar{s}s\rangle \simeq - f_\pi^2 (m_K^2-m_\pi^2)~,
\end{eqnarray}
where $f_\pi=(92.07\pm0.85)$ MeV~\cite{ParticleDataGroup:2022pth} is the pion decay constant. The complete dimension $D=6$ and $D=8$ operator are parameterized phenomenologically using the vacuum expectation values $\langle{\cal O}_6\rangle$ and $\langle{\cal O}_8\rangle$, respectively. Note that in zeroth order $\alpha_s$, i.e. neglecting running quark masses, non-perturbative dimensions do not contribute to the integral, $R(s)=\frac{1}{2\pi i}\lim_{\epsilon\to 0_{+}}\int_{s+i\epsilon}^{s-i\epsilon}{D(-\zeta)}/{\zeta}d\zeta$.
Thus in the formula presented in Eq.~(\ref{eq:nonpdef}) only the gluon and quark condensates contribute to $R(s)$ via the $\ln s$-dependence of the terms in first order $\alpha_s$.
The gluon condensate cannot be fixed theoretically. There exist experimental determinations using finite-energy sum rule techniques: a fit using the $\tau$ vector plus axial-vector hadronic width and spectral moments yields, $\left\langle(\alpha_s/\pi)GG\right\rangle=(0.001\pm0.015) {\rm GeV}^4$ \cite{ALEPH:1998rgl}, a moment analysis using $c{\bar c}$ resonances results in, $\left\langle(\alpha_s/\pi)GG\right\rangle=(0.017\pm0.004) {\rm GeV}^4$~\cite{Reinders:1984sr}, an estimation on $e^+e^-$ data gives the value of $\left\langle(\alpha_s/\pi)GG\right\rangle=(0.044^{+0.004}_{-0.021})~{\rm GeV}^4$ \cite{Bertlmann:1987ty}, a later fit on $e^+e^-$ data yields, $\langle(\alpha_s/\pi) GG\rangle=(0.037\pm0.019)~{\rm GeV}^4$ \cite{Davier:1997vd}. Due to the non-perturbative parameter $\left\langle(\alpha_s/\pi)GG\right\rangle$ fitted by different works are very different, the non-perturbative contribution will not be directly added to $R_{\rm uds}(s)$, but just provided as a theoretical error, $\pm|R_{\rm uds, NP}|_{\rm MAX}$, where the subscript ``MAX'' means the maximum of the absolute value $|R_{\rm uds, NP}|$ when $\left\langle(\alpha_s/\pi)GG\right\rangle$ varying between the upper and lower bounds for all mentioned values, i.e., $\left\langle(\alpha_s/\pi)GG\right\rangle\in[-0.014,0.056]$. Using these settings, we thus obtain a conservative estimate of the non-perturbative contribution for $R_{\rm uds}(s)$, e.g., if taking the input parameter $\alpha_s(M_Z^2)=0.1193\pm0.0028$ \cite{Baak:2014ora}, yielding $\Lambda=352^{+44}_{-41}$ MeV, we obtain $\Delta R_{\rm uds}(\sqrt{s}=5\Lambda=1.76{\rm GeV})|_{\rm NP}=\pm0.0006$, $\Delta R_{\rm uds}(\sqrt{s}=8\Lambda=2.816{\rm GeV})|_{\rm NP}=\pm0.00005$. Remember that the non-perturbative contribution decreases rapidly as the centre-of-mass $\sqrt{s}$ increases, since it is proportional to $1/s^2$.

There are thus total three theoretical errors for $R_{\rm uds}(s)$ in our calculation: $\Delta R_{\rm uds}|_{\Delta Q_*}$, $\Delta R_{\rm uds}|_{\rm UHO}$, and $\Delta R_{\rm uds}|_{\rm NP}$.
The total theoretical uncertainty is then determined by quadratically adding all the mentioned three errors.

It should be mentioned that our PMC calculation is based on the expression (\ref{eq:Rudsconv}), which is obtained in the fixed-order perturbation theory (FOPT). The PMC procedure provides a resummation of all known higher order $\{\beta_i\}$ terms, thus a further improvement for the perturbation calculation of $R(s)$. Another popular method to calculate $R(s)$ is to evaluate numerically the contour-integral, $R(s)=\frac{1}{2\pi i}\lim_{\epsilon\to 0_{+}}\int_{s+i\epsilon}^{s-i\epsilon}{D(-\zeta)}/{\zeta}d\zeta$, known as the contour-improved perturbation theory (CIPT) \cite{Pivovarov:1991rh, LeDiberder:1992jjr}. The numerical solution of the contour-integral in CIPT involves the complete (known) RGE and provides thus a resummation of all known higher order logarithmic terms. The CIPT thus can also be applied to further improving the perturbation theory for $R(s)$. The CIPT has also been widely used for the extraction of $\alpha_s$ from $\tau$ lepton decays, e.g., Refs. \cite{Baikov:2008jh, Cvetic:2010ut, Beneke:2012vb, Davier:2013sfa, Boito:2014sta, Pich:2016bdg, Ayala:2021yct, Pich:2022tca}, and the extraction of $\alpha_s$ from $e^+e^-\to {\rm hadron}$ data, see, e.g., Ref. \cite{Boito:2018yvl}. The PMC calculation will also be used for the extraction of $\alpha_s$ from $R(s)$ data in next subsection. Further comparative investigation for the extraction of $\alpha_s$ is left to future work.

\subsection{Determination of $\alpha_s$}

We adopt the PMC prediction (\ref{eq:pmcs}) as the input to fit the $R_{\rm uds}$ data in the energy range $1.84\;{\rm GeV}\sim 3.72\;{\rm GeV}$ measured by KEDR Collaboration~\cite{KEDR:2018hhr}. All the data summarized in Table $16$ of Ref.\cite{KEDR:2018hhr} are not independent but rather have point-by-point correlated effects, then the least squares (LS) estimators are determined by the minimum of $\chi^2$ function~\cite{ParticleDataGroup:2022pth},
\begin{eqnarray}\label{eq:chi2}
\chi^2(\Lambda) = \left(\textbf{e}-\textbf{t}\right)^T V^{-1}\left(\textbf{e}-\textbf{t}\right),
\end{eqnarray}
where $\textbf{e}=(R_{\rm uds}^{\rm exp.}(s_1),R_{\rm uds}^{\rm exp.}(s_2),\cdots,R_{\rm uds}^{\rm exp.}(s_{N}))$ is the column vector composed of $N=22$ experimental data, and $\textbf{t}=(R_{\rm uds}^{\rm the.}(s_1),R_{\rm uds}^{\rm the.}(s_2),\cdots,R_{\rm uds}^{\rm the.}(s_N))$ is the corresponding column vector composed of theoretical predictions. The superscript $T$ denotes the transpose. $V^{-1}$ is the inverse covariance matrix which is derived from statistical errors and systematic uncertainties taking into account the correlation matrix presented in Table $18$ of Ref.~\cite{KEDR:2018hhr}. The experimental uncertainty of the fitted parameter $\Lambda$ is determined by requiring~\cite{ParticleDataGroup:2022pth}
\begin{eqnarray}
\chi^2(\Lambda)=\chi^2_{\rm min}+1.
\end{eqnarray}
The fitting results are presented in Table \ref{tab:PMCfit}, where the first error is the experimental uncertainty. The second, third and fourth errors in $3$rd column represent contributions from the residual scale dependence $\Delta Q_*$, the UHO of the pQCD approximant $R_{\rm uds}^{(\ell)}|_{\rm PMCs}$, and the non-perturbative power correction respectively. The second errors in $4$th and $5$th columns are determined by quadratically adding all the above mentioned three components, which represent the total theoretical uncertainty.
All the components of the theoretical uncertainty are also determined by the minimum of $\chi^2$ function (\ref{eq:chi2}), but calculating the theoretical predictions $R_{\rm uds}^{\rm the.}(s_i)$ $(=R^{(\ell)}_{\rm uds}(s_i))$ in the vector $\textbf{t}=(R_{\rm uds}^{\rm the.}(s_1),R_{\rm uds}^{\rm the.}(s_2),\cdots,R_{\rm uds}^{\rm the.}(s_N))$ by adding the corresponding theoretical errors, respectively. When calculating $R_{\rm uds}^{\rm the.}(s_i)$ $(=R^{(\ell)}_{\rm uds}(s_i))$ by taking various QCD corrections ($\ell=2,3,4$) to fit the data, the running of the QCD coupling is changed accordingly, i.e, $R^{(\ell)}_{\rm uds}$ corresponding to $\ell$-loop $\alpha_s$-running. For the computation of $\alpha_s(M_Z^2)$ based on $\Lambda=\Lambda_{\overline{\rm MS}}^{(3)}$, we use the RunDec routine to firstly computing $\Lambda_{\overline{\rm MS}}^{(4)}$ and $\Lambda_{\overline{\rm MS}}^{(5)}$ and finally extract $\alpha_s(M_Z^2)$, as suggested by Ref. \cite{Herren:2017osy}.

\begin{table}[tbh]
\caption{The fitted $\Lambda$ from $R_{\rm uds}$ data below the $D\bar D$ threshold measured by KEDR collaboration~\cite{KEDR:2018hhr}. Results for $R_{\rm uds}^{\rm the.}=R_{\rm uds}^{(\ell)}|_{\rm PMCs}$ by taking various QCD corrections $(\ell=2,3,4)$ are presented. The first error is experimental, the second, third, and fourth errors of the 3rd column are from the residual scale dependence $\Delta Q_*$, UHO, and the non-perturbative contribution, respectively. The second errors of the 4th and 5th columns are total theoretical uncertainties,
which include all the mentioned three theoretical errors added in quadrature.}\label{tab:PMCfit}
\begin{center}
\begin{tabular}{ccccc}
\hline
~$R_{\rm uds}^{\rm the.}$~ & ~$\chi^2_{\rm min}/n_{\rm d.o.f.}$~ & ~$\Lambda$\,[MeV]~ & ~$\Lambda$\,[MeV]~ & ~$\alpha_s(M_Z^2)$~  \\ \hline
$R_{\rm uds}^{(2)}|_{\rm PMCs}$ & $11.0717/21$ & $333^{+138+157+70+1}_{-143-107-42-1}$ & $333^{+138+172}_{-143-115}$ & $0.1170^{+0.0082+0.0099}_{-0.0109-0.0085}$ \\ \hline
$R_{\rm uds}^{(3)}|_{\rm PMCs}$ & $10.6954/21$ & $370^{+177+80+14+2}_{-166-66-12-1}$ & $370^{+177+81}_{-166-67}$ & $0.1206^{+0.0104+0.0050}_{-0.0125-0.0046}$ \\ \hline
$R_{\rm uds}^{(4)}|_{\rm PMCs}$ & $10.5706/21$ & $406^{+207+27+4+2}_{-186-25-3-1}$ & $406^{+207+27}_{-186-25}$ & $0.1227^{+0.0117+0.0016}_{-0.0132-0.0016}$ \\
\hline
\end{tabular}
\end{center}
\end{table}

\begin{figure}[htbp]
\centering
\includegraphics[width=0.6\textwidth]{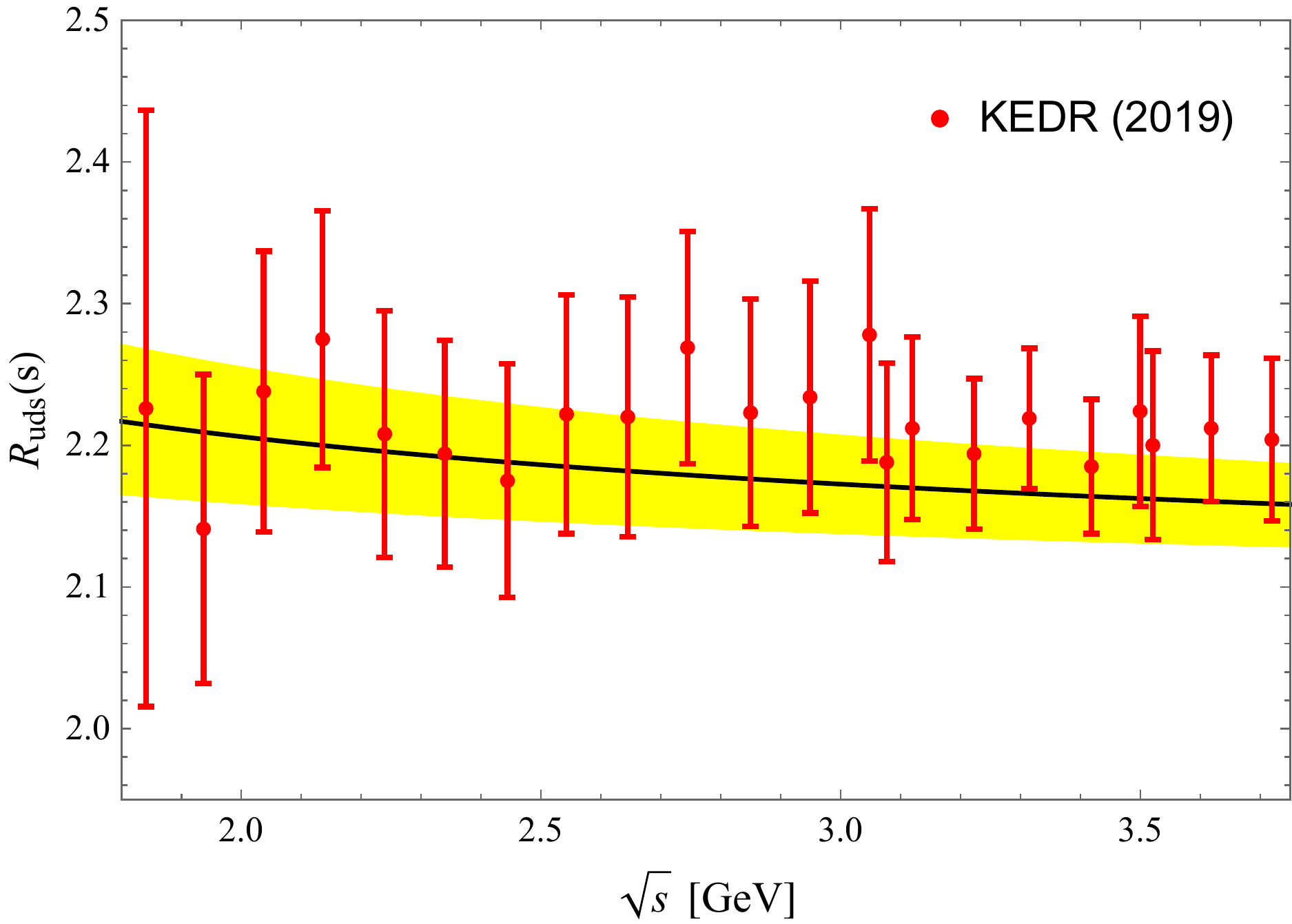}
\caption{Comparison of the fitted curve and the KEDR~\cite{KEDR:2018hhr} data with statistical and systematic errors added in quadrature. The black thin line is for $R^{(4)}_{\rm uds}(s)|_{\rm PMCs}$ and its band is plotted by taking the fitted value $\Lambda=406^{+207}_{-186}$ MeV.}
\label{fig:fit}
\end{figure}

We also present the value of $\chi^2_{\rm min}$, which represents the level of agreement between the measurements and the fitted function, and can be used for assessing the goodness-of-fit. For the $4$-loop pQCD correction $R^{(4)}_{\rm uds}(s)$, its $\chi^2_{\rm min}/n_{\rm d.o.f.}=10.5706/21\simeq 0.50$, which corresponds to $p>95\%$, indicating a good goodness-of-fit and the reasonableness of the fitted parameter $\Lambda$. A comparison of $R^{(4)}_{\rm uds}(s)|_{\rm PMCs}$ with and the KEDR data is presented in Figure \ref{fig:fit}. The resultant $\alpha_s(M_Z^2)=0.1227^{+0.0117+0.0016}_{-0.0132-0.0016}$ is consistent with the world average $\alpha_s(M_Z^2)=0.1179\pm0.0009$~\cite{ParticleDataGroup:2022pth}. Theoretical uncertainty is $\sim1.3\%$, and is negligible compared to the experimental one ($\sim10\%$). Thus the accurate theoretical prediction (\ref{eq:pmcs}) for $R_{\rm uds}(s)$ allows to extract $\alpha_s$ with high precision at the future Tau-Charm facility, such as the Super Tau-Charm Facility in China \cite{Huang:2017wbc, Peng:2020orp, Achasov:2023gey}.

\begin{figure}[htbp]
\centering
\includegraphics[width=0.6\textwidth]{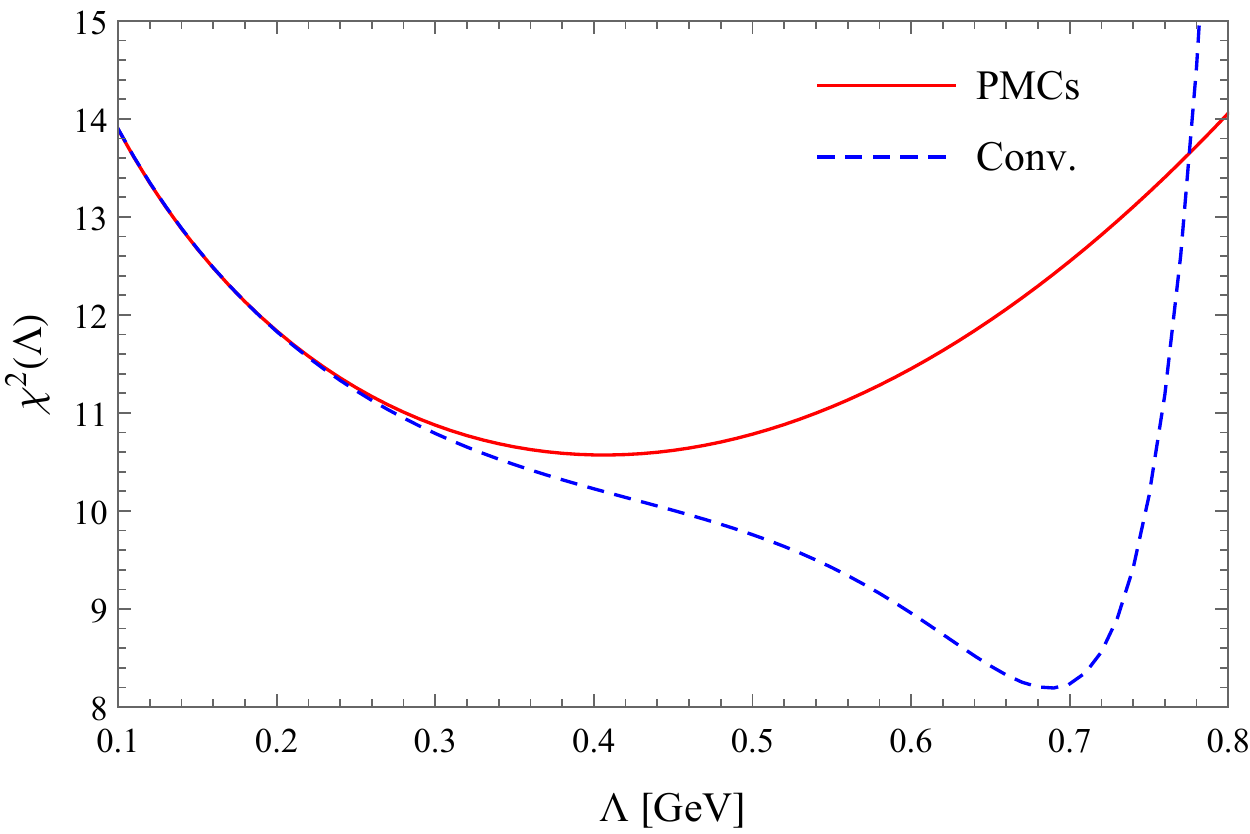}
\caption{The $\chi^2(\Lambda)$ as a function of $\Lambda$. The red thin line represents that in the definition of $\chi^2(\Lambda)$ the theoretical prediction is taking as $4$-loop PMC prediction, i.e., $R_{\rm uds}^{\rm the.}=R_{\rm uds}^{(4)}|_{\rm PMCs}$. The blue dashed line is for taking $4$-loop conventional prediction, i.e., $R_{\rm uds}^{\rm the.}=R_{\rm uds}^{(4)}|_{\rm Conv.}$.}
\label{fig:chi2}
\end{figure}

It is necessary to emphasize that when using the conventional prediction to fit $R_{\rm uds}$ data, the fitted $\Lambda$ should satisfy the self-consistent requirement $\sqrt{s}/\Lambda>{\rm exp}(\pi/2)\simeq4.81$. If using the conventional $4$-loop prediction (\ref{eq:conv}) to fit the KEDR $R_{\rm uds}$ data~\cite{KEDR:2018hhr}, one may obtain an abnormal $\chi^2(\Lambda)$ curve, see Figure \ref{fig:chi2}, and an exaggerated $\Lambda$, thus all $16$ data points below $3.3$ GeV shall violate this constraint. Such violation can be highly improved if using the $4$-loop PMCs prediction (\ref{eq:pmcs}) to fit the data, where only the first $2$ data points below $2$ GeV violate this constraint. Note that the $2$-loop and $3$-loop fit will not violate this constraint, but will have larger theoretical errors.

\subsection{Contributions to $(g-2)_\mu$ and $\alpha(M_Z^2)$}

Hadronic vacuum polarization (HVP) is not only a critical part of the Standard Model (SM) prediction for the anomalous magnetic moment of the muon, $a_\mu={(g-2)_\mu}/2$, but also a crucial ingredient for global fits to electroweak (EW) precision observables due to its contribution to the running of the fine-structure constant encoded in $\Delta\alpha_{\rm had}(q^2)$. Traditionally, the leading order HVP contribution to $a_\mu$ can be determined via the dispersion relation~\cite{Brodsky:1967sr,Lautrup:1968tdb}
\begin{equation}\label{amu_HVP}
a_\mu^{\rm HVP,LO}=\frac{\alpha(0)^2}{3\pi^2}\int_{s_{\rm th}}^{\infty} \frac{K(s)}{s}R(s) {\rm d}s,
\end{equation}
where $s_{\rm th}=m_{\pi}^2$, $\alpha(0)$ is the fine-structure constant in the Thomson limit, the kernel function $K(s)$ can be expressed analytically \cite{Brodsky:1967sr,Lautrup:1968tdb}.

The running (scale-dependent) QED coupling, $\alpha(q^2)$ is determined via,
\begin{equation}\label{eq:delAlpha}
\alpha(q^2)^{-1}=\alpha(0)^{-1}\big(1-\Delta\alpha_{\rm had}(q^2)-\Delta\alpha_{\rm lep}(q^2)\big),
\end{equation}
where the contributions to the running are separated into hadronic (had) and leptonic (lep) components. The effective QED coupling at the $Z$ boson mass, $\alpha(M_{Z}^2)$, is the least precisely known of the three fundamental electro-weak (EW) parameters of the SM (the Fermi constant $G_F$, $M_Z$ and $\alpha(M_{Z}^2)$), and its uncertainty from hadronic contributions hinders the accuracy of EW precision fits. The hadronic contributions to $\alpha(q^2)$ are determined from the dispersion relation
\begin{equation}\label{eq:Alphahad}
\Delta\alpha_{\rm had}(q^2) = -\frac{\alpha(0) q^2}{3\pi}{\rm P} \int^{\infty}_{s_{th}} \frac{R(s)}{s(s-q^2)} {\rm d}s\,,
\end{equation}
where P indicates the principal value of the integral.

\begin{figure}[htbp]
\centering
\includegraphics[width=0.6\textwidth]{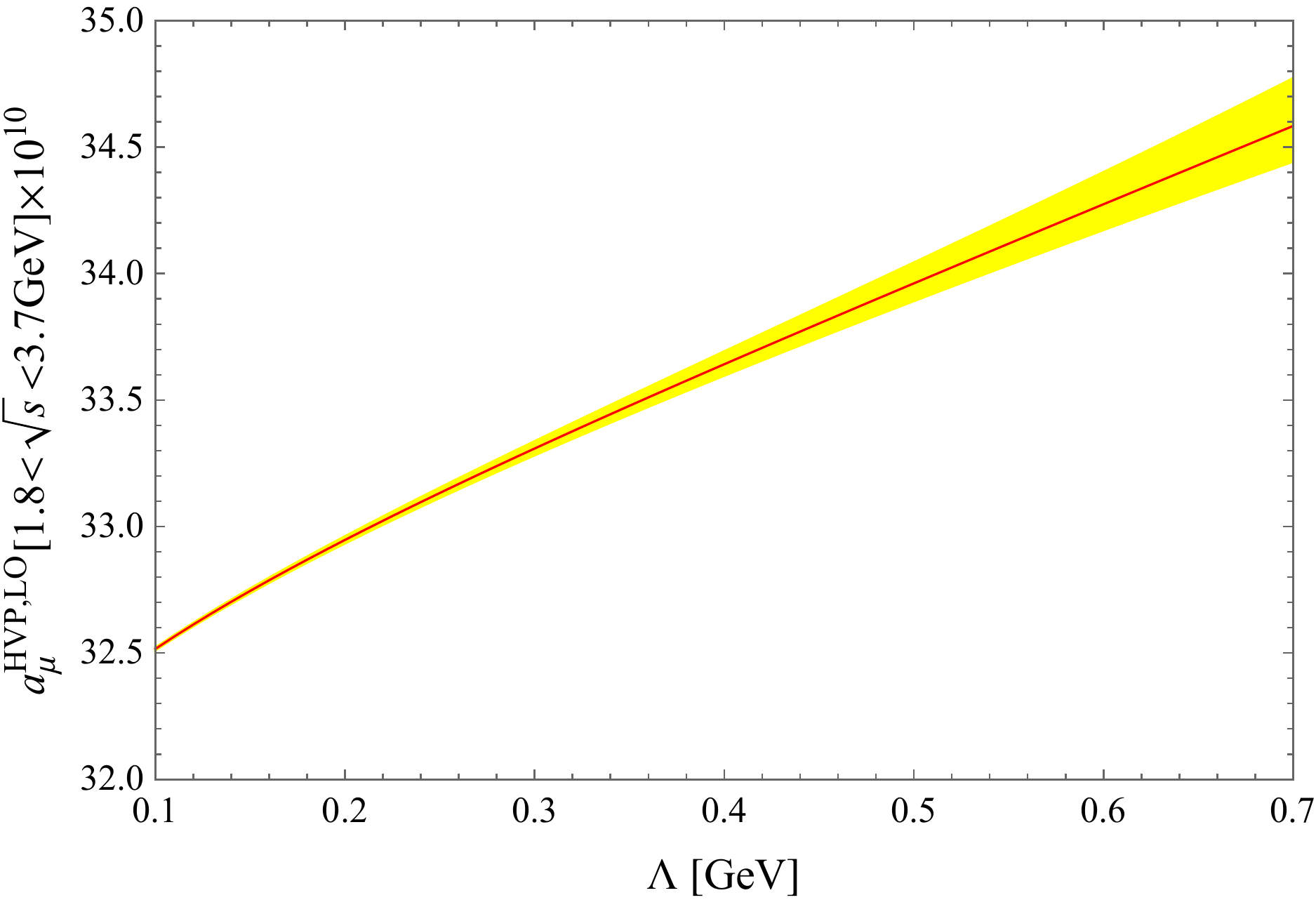}
\caption{$a_\mu^{\rm HVP,LO}[1.8\leq\sqrt{s}\leq3.7{\rm GeV}]$ as a function of $\Lambda$. The band shows the total theoretical uncertainty, including effects from the residual scale dependence $\Delta Q_*$, the UHO contribution, and the non-perturbative contribution.}
\label{fig:g2}
\end{figure}

Using the PMC prediction (\ref{eq:pmcs}), we evaluate the contribution of $R_{\rm uds}(s)$ to $a_\mu^{\rm HVP,LO}$ in energy range $1.8\sim3.7$ GeV, and present it as a function of $\Lambda$ in Fig.\ref{fig:g2}, where the band represents the theoretical uncertainty, including contributions from the residual scale dependence and the UHO of the pQCD approximant for $R_{\rm uds}(s)$ (\ref{eq:pmcs}). Such theoretical uncertainty is $\sim 0.02\%$ at $\Lambda=0.1\,{\rm GeV}$ and increases to $\sim 0.5\%$ at $\Lambda=0.7\,{\rm GeV}$. As for numerical results, if taking the same input $\alpha_s(M_Z^2)=0.1182\pm0.0012$ as the KNT18 \cite{Keshavarzi:2018mgv}, we obtain $a_\mu^{\rm HVP,LO}[1.841\leq\sqrt{s}\leq2.00{\rm GeV}]\times 10^{10}=6.38\pm 0.02$, where the total uncertainty includes effects from the $\alpha_s$ uncertainty, the residual scale dependence, the UHO contribution, and the non-perturbative contribution. This result is in good agreement with the one reported by KNT18 \cite{Keshavarzi:2018mgv}, $a_\mu^{\rm HVP,LO}[1.841\leq\sqrt{s}\leq2.00{\rm GeV}]\times 10^{10}=6.38\pm 0.11$, but with a decreased error, whose error is dominated by the variation of the renormalization scale $\mu_r$ in the range $\sqrt{s}/2<\mu_r<2\sqrt{s}$. When taking the same input as the DHMZ19 \cite{Davier:2019can}, i.e. $\alpha_s(M_Z^2)=0.1193\pm0.0028$ from the fit to $Z$ precision data \cite{Baak:2014ora}, we obtain,
\begin{equation}\label{eq:g-2}
\hspace{-1mm}a_\mu^{\rm HVP,LO}[1.8\leq\sqrt{s}\leq3.7{\rm GeV}]\times 10^{10}=33.49\pm0.15,
\end{equation}
where the error is obtained by quadratically adding the uncertainties
from the $\alpha_s$ uncertainty $(\pm0.14)$, the residual scale dependence $(\pm0.04)$, the UHO contribution $(\pm0.01)$, and the non-perturbative contribution $(\pm 0.00)$. As for the running electromagnetic coupling at $M_Z$, our prediction for the hadronic contribution from $1.8\sim3.7$ GeV range to the running of $\alpha(M_Z^2)$,
\begin{equation}\label{eq:alpha}
\Delta\alpha_{\rm had}(M_Z^2)[1.8\leq\sqrt{s}\leq3.7{\rm GeV}]\times10^{4}=24.28\pm0.10,
\end{equation}
whose uncertainty is dominated by the $\alpha_s$ uncertainty $(\pm0.10)$. The residual scale dependence $(\pm0.02)$, the UHO contribution $(\pm 0.00)$, and the non-perturbative contribution $(\pm 0.00)$ are quite small. Our present predictions (\ref{eq:g-2}) and (\ref{eq:alpha}) are in good agreement with the results reported by DHMZ19 \cite{Davier:2019can} but with decreased errors.

\section{Summary}
\label{sec:summary}

The hadronic $e^+e^-$ annihilation rate $R(s)$ is one of the most precise and theoretically safe observables involving strong interactions. The PMC provides a systematic method for solving the conventional renormalization scheme-and-scale ambiguities, and its PMC scale reflects the virtuality of the underlying QCD subprocess. By applying the PMCs, we have shown that a reliable and self-consistent analysis for $R_{\rm uds}(s)$ can be achieved. Our new calculation for $R_{\rm uds}(s)$ leads to a scale-invariant prediction, a significant stabilization of the perturbative series, and a reduction of theoretical uncertainty. It thus can provide a reliable and competitive determination for the QCD running coupling at future high-precision measurement on $R_{\rm uds}(s)$, and will help to improve the accuracy of the SM predictions for the muon magnetic anomaly $a_\mu$ as well as the QED coupling $\alpha(M_Z^2)$.

\hspace{2cm}

\noindent {\bf Acknowledgments:} We are grateful to Prof. Guang-Shun Huang, Hai-Ming Hu, Shu-Lei Zhang, Andrei Kataev, and Sergey Mikhailov for helpful discussions. This work was supported in part by the Natural Science Foundation of China under Grant No.11905056, No.12147102, No.12175025 and No.12265011.

\hspace{2cm}

\bibliographystyle{JHEP}

\bibliography{references}

\providecommand{\href}[2]{#2}\begingroup\raggedright\begin{thebibliography}{10}

\bibitem{Gross:1973id}
D.~J. Gross and F.~Wilczek, {\it {Ultraviolet Behavior of Nonabelian Gauge
  Theories}},  {\em Phys. Rev. Lett.} {\bf 30} (1973) 1343--1346.

\bibitem{Politzer:1973fx}
H.~D. Politzer, {\it {Reliable Perturbative Results for Strong Interactions?}},
   {\em Phys. Rev. Lett.} {\bf 30} (1973) 1346--1349.

\bibitem{ParticleDataGroup:2022pth}
{\bf Particle Data Group} Collaboration, R.~L. Workman et~al., {\it {Review of
  Particle Physics}},  {\em PTEP} {\bf 2022} (2022) 083C01.

\bibitem{Deur:2016tte}
A.~Deur, S.~J. Brodsky, and G.~F. de~Teramond, {\it {The QCD Running
  Coupling}},  {\em Nucl. Phys.} {\bf 90} (2016) 1,
  [\href{http://arxiv.org/abs/1604.08082}{{\tt arXiv:1604.08082}}].

\bibitem{dEnterria:2022hzv}
D.~d'Enterria et~al., {\it {The strong coupling constant: State of the art and
  the decade ahead}},  \href{http://arxiv.org/abs/2203.08271}{{\tt
  arXiv:2203.08271}}.

\bibitem{Deur:2023dzc}
A.~Deur, S.~J. Brodsky, and C.~D. Roberts, {\it {QCD Running Couplings and
  Effective Charges}},  \href{http://arxiv.org/abs/2303.00723}{{\tt
  arXiv:2303.00723}}.

\bibitem{Chetyrkin:1996ela}
K.~G. Chetyrkin, J.~H. Kuhn, and A.~Kwiatkowski, {\it {QCD corrections to the
  $e^{+} e^{-}$ cross-section and the $Z$ boson decay rate}},  {\em Phys.
  Rept.} {\bf 277} (1996) 189--281,
  [\href{http://arxiv.org/abs/hep-ph/9503396}{{\tt hep-ph/9503396}}].

\bibitem{Aoyama:2020ynm}
T.~Aoyama et~al., {\it {The anomalous magnetic moment of the muon in the
  Standard Model}},  {\em Phys. Rept.} {\bf 887} (2020) 1--166,
  [\href{http://arxiv.org/abs/2006.04822}{{\tt arXiv:2006.04822}}].

\bibitem{Jegerlehner:2017gek}
F.~Jegerlehner, {\em {The Anomalous Magnetic Moment of the Muon}}, vol.~274.
\newblock Springer, Cham, 2017.

\bibitem{BESIII:2021wib}
{\bf BESIII} Collaboration, M.~Ablikim et~al., {\it {Measurement of the Cross
  Section for $e^{+}e^{-}\to$Hadrons at Energies from 2.2324 to 3.6710~GeV}},
  {\em Phys. Rev. Lett.} {\bf 128} (2022) 062004,
  [\href{http://arxiv.org/abs/2112.11728}{{\tt arXiv:2112.11728}}].

\bibitem{KEDR:2018hhr}
{\bf KEDR} Collaboration, V.~V. Anashin et~al., {\it {Precise measurement of
  $R_{\text{uds}}$ and $R$ between 1.84 and 3.72 GeV at the KEDR detector}},
  {\em Phys. Lett. B} {\bf 788} (2019) 42--51,
  [\href{http://arxiv.org/abs/1805.06235}{{\tt arXiv:1805.06235}}].

\bibitem{BESIII:2009fln}
{\bf BESIII} Collaboration, M.~Ablikim et~al., {\it {Design and Construction of
  the BESIII Detector}},  {\em Nucl. Instrum. Meth. A} {\bf 614} (2010)
  345--399, [\href{http://arxiv.org/abs/0911.4960}{{\tt arXiv:0911.4960}}].

\bibitem{Anashin:2013twa}
V.~V. Anashin et~al., {\it {The KEDR detector}},  {\em Phys. Part. Nucl.} {\bf
  44} (2013) 657--702.

\bibitem{Davier:2005xq}
M.~Davier, A.~Hocker, and Z.~Zhang, {\it {The Physics of Hadronic Tau Decays}},
   {\em Rev. Mod. Phys.} {\bf 78} (2006) 1043--1109,
  [\href{http://arxiv.org/abs/hep-ph/0507078}{{\tt hep-ph/0507078}}].

\bibitem{Chetyrkin:1979bj}
K.~G. Chetyrkin, A.~L. Kataev, and F.~V. Tkachov, {\it {Higher Order
  Corrections to $\sigma_{\rm tot} (e^+ e^- \to Hadrons)$ in Quantum
  Chromodynamics}},  {\em Phys. Lett. B} {\bf 85} (1979) 277--279.

\bibitem{Gorishnii:1990vf}
S.~G. Gorishnii, A.~L. Kataev, and S.~A. Larin, {\it {The
  $O(\alpha^{3}_{s})$-corrections to $\sigma_{tot}(e^{+}e^{-}\rightarrow
  hadrons)$ and $\Gamma(\tau^{-} \rightarrow \nu_{\tau} + hadrons)$ in QCD}},
  {\em Phys. Lett. B} {\bf 259} (1991) 144--150.

\bibitem{Surguladze:1990tg}
L.~R. Surguladze and M.~A. Samuel, {\it {Total hadronic cross-section in $e^+
  e^-$ annihilation at the four loop level of perturbative QCD}},  {\em Phys.
  Rev. Lett.} {\bf 66} (1991) 560--563. [Erratum: Phys.Rev.Lett. 66, 2416
  (1991)].

\bibitem{Baikov:2008jh}
P.~A. Baikov, K.~G. Chetyrkin, and J.~H. Kuhn, {\it {Order $\alpha^4(s)$ QCD
  Corrections to Z and tau Decays}},  {\em Phys. Rev. Lett.} {\bf 101} (2008)
  012002, [\href{http://arxiv.org/abs/0801.1821}{{\tt arXiv:0801.1821}}].

\bibitem{Baikov:2010je}
P.~A. Baikov, K.~G. Chetyrkin, and J.~H. Kuhn, {\it {Adler Function, Bjorken
  Sum Rule, and the Crewther Relation to Order $\alpha^4_s$ in a General Gauge
  Theory}},  {\em Phys. Rev. Lett.} {\bf 104} (2010) 132004,
  [\href{http://arxiv.org/abs/1001.3606}{{\tt arXiv:1001.3606}}].

\bibitem{Baikov:2012er}
P.~A. Baikov, K.~G. Chetyrkin, J.~H. Kuhn, and J.~Rittinger, {\it {Complete
  ${\cal O}(\alpha_s^4)$ QCD Corrections to Hadronic $Z$-Decays}},  {\em Phys.
  Rev. Lett.} {\bf 108} (2012) 222003,
  [\href{http://arxiv.org/abs/1201.5804}{{\tt arXiv:1201.5804}}].

\bibitem{Baikov:2012zm}
P.~A. Baikov, K.~G. Chetyrkin, J.~H. Kuhn, and J.~Rittinger, {\it {Vector
  Correlator in Massless QCD at Order $\mathcal{O}(\alpha^4_s)$ and the QED
  beta-function at Five Loop}},  {\em JHEP} {\bf 07} (2012) 017,
  [\href{http://arxiv.org/abs/1206.1284}{{\tt arXiv:1206.1284}}].

\bibitem{Baikov:2012zn}
P.~A. Baikov, K.~G. Chetyrkin, J.~H. Kuhn, and J.~Rittinger, {\it {Adler
  Function, Sum Rules and Crewther Relation of Order $\mathcal{O}(\alpha^4_s)$:
  the Singlet Case}},  {\em Phys. Lett. B} {\bf 714} (2012) 62--65,
  [\href{http://arxiv.org/abs/1206.1288}{{\tt arXiv:1206.1288}}].

\bibitem{Chetyrkin:1990kr}
K.~G. Chetyrkin and J.~H. Kuhn, {\it {Mass corrections to the Z decay rate}},
  {\em Phys. Lett. B} {\bf 248} (1990) 359--364.

\bibitem{Chetyrkin:1994ex}
K.~G. Chetyrkin and J.~H. Kuhn, {\it {Quartic mass corrections to $R_{\rm
  had}$}},  {\em Nucl. Phys. B} {\bf 432} (1994) 337--350,
  [\href{http://arxiv.org/abs/hep-ph/9406299}{{\tt hep-ph/9406299}}].

\bibitem{Chetyrkin:1995ii}
K.~G. Chetyrkin, J.~H. Kuhn, and M.~Steinhauser, {\it {Heavy quark vacuum
  polarization to three loops}},  {\em Phys. Lett. B} {\bf 371} (1996) 93--98,
  [\href{http://arxiv.org/abs/hep-ph/9511430}{{\tt hep-ph/9511430}}].

\bibitem{Chetyrkin:1996cf}
K.~G. Chetyrkin, J.~H. Kuhn, and M.~Steinhauser, {\it {Three loop polarization
  function and ${\cal O} (\alpha_s^2)$ corrections to the production of heavy
  quarks}},  {\em Nucl. Phys. B} {\bf 482} (1996) 213--240,
  [\href{http://arxiv.org/abs/hep-ph/9606230}{{\tt hep-ph/9606230}}].

\bibitem{Chetyrkin:2000zk}
K.~G. Chetyrkin, R.~V. Harlander, and J.~H. Kuhn, {\it {Quartic mass
  corrections to $R_{had}$ at $\mathcal O(\alpha^3_s)$}},  {\em Nucl. Phys. B}
  {\bf 586} (2000) 56--72, [\href{http://arxiv.org/abs/hep-ph/0005139}{{\tt
  hep-ph/0005139}}]. [Erratum: Nucl.Phys.B 634, 413--414 (2002)].

\bibitem{Harlander:2002ur}
R.~V. Harlander and M.~Steinhauser, {\it {rhad: A Program for the evaluation of
  the hadronic R ratio in the perturbative regime of QCD}},  {\em Comput. Phys.
  Commun.} {\bf 153} (2003) 244--274,
  [\href{http://arxiv.org/abs/hep-ph/0212294}{{\tt hep-ph/0212294}}].

\bibitem{Kiyo:2009gb}
Y.~Kiyo, A.~Maier, P.~Maierhofer, and P.~Marquard, {\it {Reconstruction of
  heavy quark current correlators at $O(\alpha_s^3)$}},  {\em Nucl. Phys. B}
  {\bf 823} (2009) 269--287, [\href{http://arxiv.org/abs/0907.2120}{{\tt
  arXiv:0907.2120}}].

\bibitem{Czarnecki:1996ei}
A.~Czarnecki and J.~H. Kuhn, {\it {Nonfactorizable QCD and electroweak
  corrections to the hadronic Z boson decay rate}},  {\em Phys. Rev. Lett.}
  {\bf 77} (1996) 3955--3958, [\href{http://arxiv.org/abs/hep-ph/9608366}{{\tt
  hep-ph/9608366}}].

\bibitem{Harlander:1998cmq}
R.~Harlander, T.~Seidensticker, and M.~Steinhauser, {\it {Complete corrections
  of Order ${\cal O}(\alpha\alpha_s)$ to the decay of the Z boson into bottom
  quarks}},  {\em Phys. Lett. B} {\bf 426} (1998) 125--132,
  [\href{http://arxiv.org/abs/hep-ph/9712228}{{\tt hep-ph/9712228}}].

\bibitem{Kuhn:1998ze}
J.~H. Kuhn and M.~Steinhauser, {\it {A Theory driven analysis of the effective
  QED coupling at $M_Z$}},  {\em Phys. Lett. B} {\bf 437} (1998) 425--431,
  [\href{http://arxiv.org/abs/hep-ph/9802241}{{\tt hep-ph/9802241}}].

\bibitem{Martin:1999bp}
A.~D. Martin, J.~Outhwaite, and M.~G. Ryskin, {\it {The R ratio in $e^+ e^-$,
  the determination of $\alpha (M^2_Z)$ and a possible nonperturbative gluonic
  contribution}},  {\em J. Phys. G} {\bf 26} (2000) 600--606,
  [\href{http://arxiv.org/abs/hep-ph/9912252}{{\tt hep-ph/9912252}}].

\bibitem{Brodsky:2011ta}
S.~J. Brodsky and X.-G. Wu, {\it {Scale Setting Using the Extended
  Renormalization Group and the Principle of Maximum Conformality: the QCD
  Coupling Constant at Four Loops}},  {\em Phys. Rev. D} {\bf 85} (2012)
  034038, [\href{http://arxiv.org/abs/1111.6175}{{\tt arXiv:1111.6175}}].
  [Erratum: Phys.Rev.D 86, 079903 (2012)].

\bibitem{Brodsky:2011ig}
S.~J. Brodsky and L.~Di~Giustino, {\it {Setting the Renormalization Scale in
  QCD: The Principle of Maximum Conformality}},  {\em Phys. Rev. D} {\bf 86}
  (2012) 085026, [\href{http://arxiv.org/abs/1107.0338}{{\tt
  arXiv:1107.0338}}].

\bibitem{Mojaza:2012mf}
M.~Mojaza, S.~J. Brodsky, and X.-G. Wu, {\it {Systematic All-Orders Method to
  Eliminate Renormalization-Scale and Scheme Ambiguities in Perturbative QCD}},
   {\em Phys. Rev. Lett.} {\bf 110} (2013) 192001,
  [\href{http://arxiv.org/abs/1212.0049}{{\tt arXiv:1212.0049}}].

\bibitem{Brodsky:2012rj}
S.~J. Brodsky and X.-G. Wu, {\it {Eliminating the Renormalization Scale
  Ambiguity for Top-Pair Production Using the Principle of Maximum
  Conformality}},  {\em Phys. Rev. Lett.} {\bf 109} (2012) 042002,
  [\href{http://arxiv.org/abs/1203.5312}{{\tt arXiv:1203.5312}}].

\bibitem{Brodsky:2013vpa}
S.~J. Brodsky, M.~Mojaza, and X.-G. Wu, {\it {Systematic Scale-Setting to All
  Orders: The Principle of Maximum Conformality and Commensurate Scale
  Relations}},  {\em Phys. Rev. D} {\bf 89} (2014) 014027,
  [\href{http://arxiv.org/abs/1304.4631}{{\tt arXiv:1304.4631}}].

\bibitem{Shen:2017pdu}
J.-M. Shen, X.-G. Wu, B.-L. Du, and S.~J. Brodsky, {\it {Novel All-Orders
  Single-Scale Approach to QCD Renormalization Scale-Setting}},  {\em Phys.
  Rev. D} {\bf 95} (2017) 094006, [\href{http://arxiv.org/abs/1701.08245}{{\tt
  arXiv:1701.08245}}].

\bibitem{Yan:2022foz}
J.~Yan, Z.-F. Wu, J.-M. Shen, and X.-G. Wu, {\it {Precise perturbative
  predictions from fixed-order calculations}},  {\em J. Phys. G} {\bf 50}
  (2023) 045001, [\href{http://arxiv.org/abs/2209.13364}{{\tt
  arXiv:2209.13364}}].

\bibitem{Wu:2018cmb}
X.-G. Wu, J.-M. Shen, B.-L. Du, and S.~J. Brodsky, {\it {Novel demonstration of
  the renormalization group invariance of the fixed-order predictions using the
  principle of maximum conformality and the $C$-scheme coupling}},  {\em Phys.
  Rev. D} {\bf 97} (2018) 094030, [\href{http://arxiv.org/abs/1802.09154}{{\tt
  arXiv:1802.09154}}].

\bibitem{Wu:2019mky}
X.-G. Wu, J.-M. Shen, B.-L. Du, X.-D. Huang, S.-Q. Wang, and S.~J. Brodsky,
  {\it {The QCD renormalization group equation and the elimination of
  fixed-order scheme-and-scale ambiguities using the principle of maximum
  conformality}},  {\em Prog. Part. Nucl. Phys.} {\bf 108} (2019) 103706,
  [\href{http://arxiv.org/abs/1903.12177}{{\tt arXiv:1903.12177}}].

\bibitem{StueckelbergdeBreidenbach:1952pwl}
E.~C.~G. Stueckelberg~de Breidenbach and A.~Petermann, {\it {Normalization of
  constants in the quanta theory}},  {\em Helv. Phys. Acta} {\bf 26} (1953)
  499--520.

\bibitem{MR0073481}
N.~N. Bogolyubov and D.~V. Shirkov, {\it Application of the renormalization
  group to improvement of formulas in perturbation theory},  {\em Dokl. Akad.
  Nauk SSSR (N.S.)} {\bf 103} (1955) 391--394.

\bibitem{Peterman:1978tb}
A.~Peterman, {\it {Renormalization Group and the Deep Structure of the
  Proton}},  {\em Phys. Rept.} {\bf 53} (1979) 157.

\bibitem{Wu:2014iba}
X.-G. Wu, Y.~Ma, S.-Q. Wang, H.-B. Fu, H.-H. Ma, S.~J. Brodsky, and M.~Mojaza,
  {\it {Renormalization Group Invariance and Optimal QCD Renormalization
  Scale-Setting}},  {\em Rept. Prog. Phys.} {\bf 78} (2015) 126201,
  [\href{http://arxiv.org/abs/1405.3196}{{\tt arXiv:1405.3196}}].

\bibitem{Brodsky:2012ms}
S.~J. Brodsky and X.-G. Wu, {\it {Self-Consistency Requirements of the
  Renormalization Group for Setting the Renormalization Scale}},  {\em Phys.
  Rev. D} {\bf 86} (2012) 054018, [\href{http://arxiv.org/abs/1208.0700}{{\tt
  arXiv:1208.0700}}].

\bibitem{Wu:2013ei}
X.-G. Wu, S.~J. Brodsky, and M.~Mojaza, {\it {The Renormalization Scale-Setting
  Problem in QCD}},  {\em Prog. Part. Nucl. Phys.} {\bf 72} (2013) 44--98,
  [\href{http://arxiv.org/abs/1302.0599}{{\tt arXiv:1302.0599}}].

\bibitem{Zheng:2013uja}
X.-C. Zheng, X.-G. Wu, S.-Q. Wang, J.-M. Shen, and Q.-L. Zhang, {\it
  {Reanalysis of the BFKL Pomeron at the next-to-leading logarithmic
  accuracy}},  {\em JHEP} {\bf 10} (2013) 117,
  [\href{http://arxiv.org/abs/1308.2381}{{\tt arXiv:1308.2381}}].

\bibitem{Gell-Mann:1954yli}
M.~Gell-Mann and F.~E. Low, {\it {Quantum electrodynamics at small distances}},
   {\em Phys. Rev.} {\bf 95} (1954) 1300--1312.

\bibitem{Brodsky:1982gc}
S.~J. Brodsky, G.~P. Lepage, and P.~B. Mackenzie, {\it {On the Elimination of
  Scale Ambiguities in Perturbative Quantum Chromodynamics}},  {\em Phys. Rev.
  D} {\bf 28} (1983) 228.

\bibitem{Adler:1974gd}
S.~L. Adler, {\it {Some Simple Vacuum Polarization Phenomenology: $e^+ e^- \to$
  Hadrons: The $\mu$ - Mesic Atom x-Ray Discrepancy and $g_{\mu}-2$}},  {\em
  Phys. Rev. D} {\bf 10} (1974) 3714.

\bibitem{Bi:2015wea}
H.-Y. Bi, X.-G. Wu, Y.~Ma, H.-H. Ma, S.~J. Brodsky, and M.~Mojaza, {\it
  {Degeneracy Relations in QCD and the Equivalence of Two Systematic All-Orders
  Methods for Setting the Renormalization Scale}},  {\em Phys. Lett. B} {\bf
  748} (2015) 13--18, [\href{http://arxiv.org/abs/1505.04958}{{\tt
  arXiv:1505.04958}}].

\bibitem{Kataev:1995vh}
A.~L. Kataev and V.~V. Starshenko, {\it {Estimates of the higher order QCD
  corrections to $R(s)$, $R(\tau)$ and deep inelastic scattering sum rules}},
  {\em Mod. Phys. Lett. A} {\bf 10} (1995) 235--250,
  [\href{http://arxiv.org/abs/hep-ph/9502348}{{\tt hep-ph/9502348}}].

\bibitem{Shirkov:2000qv}
D.~V. Shirkov, {\it {Analytic perturbation theory for QCD observables}},  {\em
  Theor. Math. Phys.} {\bf 127} (2001) 409--423,
  [\href{http://arxiv.org/abs/hep-ph/0012283}{{\tt hep-ph/0012283}}].

\bibitem{Prosperi:2006hx}
G.~M. Prosperi, M.~Raciti, and C.~Simolo, {\it {On the running coupling
  constant in QCD}},  {\em Prog. Part. Nucl. Phys.} {\bf 58} (2007) 387--438,
  [\href{http://arxiv.org/abs/hep-ph/0607209}{{\tt hep-ph/0607209}}].

\bibitem{Nesterenko:2017wpb}
A.~V. Nesterenko, {\it {Electron\textendash{}positron annihilation into hadrons
  at the higher-loop levels}},  {\em Eur. Phys. J. C} {\bf 77} (2017) 844,
  [\href{http://arxiv.org/abs/1707.00668}{{\tt arXiv:1707.00668}}].

\bibitem{Nesterenko:2019rag}
A.~V. Nesterenko, {\it {Explicit form of the R-ratio of
  electron\textendash{}positron annihilation into hadrons}},  {\em J. Phys. G}
  {\bf 46} (2019) 115006, [\href{http://arxiv.org/abs/1902.06504}{{\tt
  arXiv:1902.06504}}].

\bibitem{Nesterenko:2020nol}
A.~V. Nesterenko, {\it {Recurrent form of the renormalization group relations
  for the higher-order hadronic vacuum polarization function perturbative
  expansion coefficients}},  {\em J. Phys. G} {\bf 47} (2020) 105001,
  [\href{http://arxiv.org/abs/2004.00609}{{\tt arXiv:2004.00609}}].

\bibitem{Cacciari:2011ze}
M.~Cacciari and N.~Houdeau, {\it {Meaningful characterisation of perturbative
  theoretical uncertainties}},  {\em JHEP} {\bf 09} (2011) 039,
  [\href{http://arxiv.org/abs/1105.5152}{{\tt arXiv:1105.5152}}].

\bibitem{Shen:2022nyr}
J.-M. Shen, Z.-J. Zhou, S.-Q. Wang, J.~Yan, Z.-F. Wu, X.-G. Wu, and S.~J.
  Brodsky, {\it {Extending the predictive power of perturbative QCD using the
  principle of maximum conformality and the Bayesian analysis}},  {\em Eur.
  Phys. J. C} {\bf 83} (2023) 326, [\href{http://arxiv.org/abs/2209.03546}{{\tt
  arXiv:2209.03546}}].

\bibitem{Bonvini:2020xeo}
M.~Bonvini, {\it {Probabilistic definition of the perturbative theoretical
  uncertainty from missing higher orders}},  {\em Eur. Phys. J. C} {\bf 80}
  (2020) 989, [\href{http://arxiv.org/abs/2006.16293}{{\tt arXiv:2006.16293}}].

\bibitem{Duhr:2021mfd}
C.~Duhr, A.~Huss, A.~Mazeliauskas, and R.~Szafron, {\it {An analysis of
  Bayesian estimates for missing higher orders in perturbative calculations}},
  {\em JHEP} {\bf 09} (2021) 122, [\href{http://arxiv.org/abs/2106.04585}{{\tt
  arXiv:2106.04585}}].

\bibitem{Braaten:1991qm}
E.~Braaten, S.~Narison, and A.~Pich, {\it {QCD analysis of the tau hadronic
  width}},  {\em Nucl. Phys. B} {\bf 373} (1992) 581--612.

\bibitem{Dokshitzer:1995qm}
Y.~L. Dokshitzer, G.~Marchesini, and B.~R. Webber, {\it {Dispersive approach to
  power behaved contributions in QCD hard processes}},  {\em Nucl. Phys. B}
  {\bf 469} (1996) 93--142, [\href{http://arxiv.org/abs/hep-ph/9512336}{{\tt
  hep-ph/9512336}}].

\bibitem{Davier:1997vd}
M.~Davier and A.~Hocker, {\it {Improved determination of $\alpha(M_Z^2)$ and
  the anomalous magnetic moment of the muon}},  {\em Phys. Lett. B} {\bf 419}
  (1998) 419--431, [\href{http://arxiv.org/abs/hep-ph/9801361}{{\tt
  hep-ph/9801361}}].

\bibitem{Gell-Mann:1960mvl}
M.~Gell-Mann and M.~Levy, {\it {The axial vector current in beta decay}},  {\em
  Nuovo Cim.} {\bf 16} (1960) 705.

\bibitem{Nambu:1960xd}
Y.~Nambu, {\it {Axial vector current conservation in weak interactions}},  {\em
  Phys. Rev. Lett.} {\bf 4} (1960) 380--382.

\bibitem{Davier:1998si}
M.~Davier and A.~Hocker, {\it {New results on the hadronic contributions to
  $\alpha(M_Z^2)$ and to $(g-2)_\mu$}},  {\em Phys. Lett. B} {\bf 435} (1998)
  427--440, [\href{http://arxiv.org/abs/hep-ph/9805470}{{\tt hep-ph/9805470}}].

\bibitem{ALEPH:1998rgl}
{\bf ALEPH} Collaboration, R.~Barate et~al., {\it {Measurement of the spectral
  functions of axial - vector hadronic tau decays and determination of
  $\alpha_s(M^2_\tau)$}},  {\em Eur. Phys. J. C} {\bf 4} (1998) 409--431.

\bibitem{Reinders:1984sr}
L.~J. Reinders, H.~Rubinstein, and S.~Yazaki, {\it {Hadron Properties from QCD
  Sum Rules}},  {\em Phys. Rept.} {\bf 127} (1985) 1.

\bibitem{Bertlmann:1987ty}
R.~A. Bertlmann, C.~A. Dominguez, M.~Loewe, M.~Perrottet, and E.~de~Rafael,
  {\it {Determination of the Gluon Condensate and the Four Quark Condensate via
  FESR}},  {\em Z. Phys. C} {\bf 39} (1988) 231.

\bibitem{Baak:2014ora}
{\bf Gfitter Group} Collaboration, M.~Baak, J.~C\'uth, J.~Haller, A.~Hoecker,
  R.~Kogler, K.~M\"onig, M.~Schott, and J.~Stelzer, {\it {The global
  electroweak fit at NNLO and prospects for the LHC and ILC}},  {\em Eur. Phys.
  J. C} {\bf 74} (2014) 3046, [\href{http://arxiv.org/abs/1407.3792}{{\tt
  arXiv:1407.3792}}].

\bibitem{Pivovarov:1991rh}
A.~A. Pivovarov, {\it {Renormalization group analysis of the tau lepton decay
  within QCD}},  {\em Sov. J. Nucl. Phys.} {\bf 54} (1991) 676--678,
  [\href{http://arxiv.org/abs/hep-ph/0302003}{{\tt hep-ph/0302003}}].

\bibitem{LeDiberder:1992jjr}
F.~Le~Diberder and A.~Pich, {\it {The perturbative QCD prediction to $R_\tau$
  revisited}},  {\em Phys. Lett. B} {\bf 286} (1992) 147--152.

\bibitem{Cvetic:2010ut}
G.~Cvetic, M.~Loewe, C.~Martinez, and C.~Valenzuela, {\it {Modified
  Contour-Improved Perturbation Theory}},  {\em Phys. Rev. D} {\bf 82} (2010)
  093007, [\href{http://arxiv.org/abs/1005.4444}{{\tt arXiv:1005.4444}}].

\bibitem{Beneke:2012vb}
M.~Beneke, D.~Boito, and M.~Jamin, {\it {Perturbative expansion of $\tau$
  hadronic spectral function moments and $\alpha_s$ extractions}},  {\em JHEP}
  {\bf 01} (2013) 125, [\href{http://arxiv.org/abs/1210.8038}{{\tt
  arXiv:1210.8038}}].

\bibitem{Davier:2013sfa}
M.~Davier, A.~H\"ocker, B.~Malaescu, C.-Z. Yuan, and Z.~Zhang, {\it {Update of
  the ALEPH non-strange spectral functions from hadronic $\tau$ decays}},  {\em
  Eur. Phys. J. C} {\bf 74} (2014) 2803,
  [\href{http://arxiv.org/abs/1312.1501}{{\tt arXiv:1312.1501}}].

\bibitem{Boito:2014sta}
D.~Boito, M.~Golterman, K.~Maltman, J.~Osborne, and S.~Peris, {\it {Strong
  coupling from the revised ALEPH data for hadronic $\tau$ decays}},  {\em
  Phys. Rev. D} {\bf 91} (2015) 034003,
  [\href{http://arxiv.org/abs/1410.3528}{{\tt arXiv:1410.3528}}].

\bibitem{Pich:2016bdg}
A.~Pich and A.~Rodr\'\i{}guez-S\'anchez, {\it {Determination of the QCD
  coupling from ALEPH $\tau$ decay data}},  {\em Phys. Rev. D} {\bf 94} (2016)
  034027, [\href{http://arxiv.org/abs/1605.06830}{{\tt arXiv:1605.06830}}].

\bibitem{Ayala:2021yct}
C.~Ayala, G.~Cvetic, and D.~Teca, {\it {Using improved operator product
  expansion in Borel\textendash{}Laplace sum rules with ALEPH $\tau $ decay
  data, and determination of pQCD coupling}},  {\em Eur. Phys. J. C} {\bf 82}
  (2022) 362, [\href{http://arxiv.org/abs/2112.01992}{{\tt arXiv:2112.01992}}].

\bibitem{Pich:2022tca}
A.~Pich and A.~Rodr\'\i{}guez-S\'anchez, {\it {Violations of quark-hadron
  duality in low-energy determinations of \ensuremath{\alpha}$_{s}$}},  {\em
  JHEP} {\bf 07} (2022) 145, [\href{http://arxiv.org/abs/2205.07587}{{\tt
  arXiv:2205.07587}}].

\bibitem{Boito:2018yvl}
D.~Boito, M.~Golterman, A.~Keshavarzi, K.~Maltman, D.~Nomura, S.~Peris, and
  T.~Teubner, {\it {Strong coupling from $e^+e^-\to$ hadrons below charm}},
  {\em Phys. Rev. D} {\bf 98} (2018) 074030,
  [\href{http://arxiv.org/abs/1805.08176}{{\tt arXiv:1805.08176}}].

\bibitem{Herren:2017osy}
F.~Herren and M.~Steinhauser, {\it {Version 3 of RunDec and CRunDec}},  {\em
  Comput. Phys. Commun.} {\bf 224} (2018) 333--345,
  [\href{http://arxiv.org/abs/1703.03751}{{\tt arXiv:1703.03751}}].

\bibitem{Huang:2017wbc}
G.~Huang et~al., {\it {Physics on the high intensive electron position
  accelerator at $2\sim 7$ GeV (in Chinese)}},  {\em Chin. Sci. Bull.} {\bf 62}
  (2017) 1226--1232.

\bibitem{Peng:2020orp}
H.~P. Peng, Y.~H. Zheng, and X.~R. Zhou, {\it {Super Tau-Charm Facility of
  China}},  {\em Physics} {\bf 49} (2020) 513--524.

\bibitem{Achasov:2023gey}
M.~Achasov et~al., {\it {STCF Conceptual Design Report: Volume I - Physics \&
  Detector}},  \href{http://arxiv.org/abs/2303.15790}{{\tt arXiv:2303.15790}}.

\bibitem{Brodsky:1967sr}
S.~J. Brodsky and E.~De~Rafael, {\it {Suggested boson\textendash{}lepton pair
  couplings and the anomalous magnetic moment of the muon}},  {\em Phys. Rev.}
  {\bf 168} (1968) 1620--1622.

\bibitem{Lautrup:1968tdb}
B.~E. Lautrup and E.~De~Rafael, {\it {Calculation of the sixth-order
  contribution from the fourth-order vacuum polarization to the difference of
  the anomalous magnetic moments of muon and electron}},  {\em Phys. Rev.} {\bf
  174} (1968) 1835--1842.

\bibitem{Keshavarzi:2018mgv}
A.~Keshavarzi, D.~Nomura, and T.~Teubner, {\it {Muon $g-2$ and $\alpha(M_Z^2)$:
  a new data-based analysis}},  {\em Phys. Rev. D} {\bf 97} (2018) 114025,
  [\href{http://arxiv.org/abs/1802.02995}{{\tt arXiv:1802.02995}}].

\bibitem{Davier:2019can}
M.~Davier, A.~Hoecker, B.~Malaescu, and Z.~Zhang, {\it {A new evaluation of the
  hadronic vacuum polarisation contributions to the muon anomalous magnetic
  moment and to $\alpha(m_Z^2)$}},  {\em Eur. Phys. J. C} {\bf 80} (2020) 241,
  [\href{http://arxiv.org/abs/1908.00921}{{\tt arXiv:1908.00921}}]. [Erratum:
  Eur.Phys.J.C 80, 410 (2020)].

\end{thebibliography}\endgroup

\end{document}